\documentclass[11pt]{article}
\usepackage{amssymb}

\usepackage{graphicx}
\usepackage[english]{babel}
\usepackage{amsmath}
\usepackage{alltt}
\usepackage{color}

\newtheorem{theorem}{Theorem}[section]

\newtheorem{summary}[theorem]{Summary}

\input{tcilatex}

\begin{document}

\title{A comparison of three turbulence models with an application to the West Pacific Warm Pool}
\author{ {\small AC.\ BENNIS}\thanks{%
IRMAR, Universit\'{e} de Rennes 1, Campus de Beaulieu, 35042 Rennes Cedex,
France}\ ,\thinspace {\small M.\ GOMEZ MARMOL} \thanks{%
Departamento de Ecuaciones Diferenciales y An\'{a}lisis Numerico,
Universidad de Sevilla.\ C/Tarfia, s/n.41080, Sevilla, Spain}\ ,\thinspace
{\small R.\ LEWANDOWSKI} \thanks{%
IRMAR, Universit\'{e} de Rennes 1, Campus de Beaulieu, 35042 Rennes Cedex,
France} , \thinspace {\small T.\ CHACON REBOLLO} \thanks{%
Departamento de Ecuaciones Diferenciales y An\'{a}lisis Numerico,
Universidad de Sevilla.\ C/Tarfia, s/n.41080, Sevilla, Spain}\,
 {\small F.\ BROSSIER}\thanks{%
IRMAR, Universit\'{e} de Rennes 1, Campus de Beaulieu, 35042 Rennes Cedex,
France}\ }
\date{}
\maketitle

\begin{abstract}
In this work, we compare three turbulence models used to
parameterize the oceanic boundary layer. These three models depend on
the bulk Richardson number, which is coherent with the studied
region, the West Pacific Warm Pool, because of the large mean shear
associated with the equatorial undercurrent. One of these models,
called R224, is new and the others are Pacanowski and
Philander's model (R213 model) and Gent's model (R23 model). The
numerical implementation is based on a non-conservative numerical
scheme. The following (three criteria) are used to compare the
models: the surface current intensity, the pycnocline's form and
the mixed layer depth. We initialize the code with realistic velocity and density profiles
thanks the TOGA-TAO array (McPhaden, 1995, \cite{Mc95}). In case of
static instability zone on the initial density profile, only the
R224 model gives realistic results. Afterwards, we study a mixed layer induced by the 
wind stress. In this case, the R224 results and the Pacanowski and Philander's results are similar.
Furthermore, we simulate a long time case. We obtain a linear solution for all models
that is in agreement with Bennis and al \cite{Be07}.

\end{abstract}

\begin{summary}
\bigskip
\end{summary}

\section{Introduction}

The presence of an homogeneous layer near the surface of the ocean
has been observed since a long time. This layer presents almost
constant profiles of temperature and salinity. We distinguish the
mixing layer from the mixed layer (Brainerd and Gregg, 1995,
\cite{Br95}). The mixing layer is actively mixed by surface fluxes.
It is a depth zone where the turbulence is strong. The mixed layer
is a maximum depth zone which has been mixed in the
recent past by surface fluxes. This layer includes the mixing layer. In this work, we
study the mixed layer. The bottom of the mixed layer corresponds
either to the top of the thermocline, zone of large gradients of
temperature, or to the top of the zone where haline stratification
is observed (Vialard and Delecluse, 1998, \cite{Vi98}). Some
attempts to describe this phenomenon can be found for example in
Defant \cite{De36} or in Lewandowski \cite{Lew97}. The effect of the
wind-stress acting on the sea-surface is considered to be the main
forcing of this boundary layer.\ Observations in situ were completed
by laboratory experiments (Deardoff, 1969, \cite{Dea69}) and more
recently by numerical modelizations of the mixed layer. A
historical of these different approaches can be found in K\"{a}se
(\cite{Ka98}). K\"{a}se quoted the model of Kraus and Turner (1967,
\cite{Kr67}) as the first applied mixed layer model which is a bulk one. 
There are also first order models as the
Pacanowski and Philander model (called PP model, 1981, \cite{Pa81})
and the Large and Gent model (called KPP model, 1994, \cite{La94}).
The second order models have been developed by Mellor and Yamada
(called MY model, 1982, \cite{Me82}) and Gaspar and al (1990,
\cite{Ga90}). In this work, we focus our attention on the first
order models. According to observations, the thickness of the mixed
layer can vary between ten meters and a few hundred of meters,
depending on the latitude (Boyer de Mont\'egut and al, 2004,
\cite{Mo04}). The mixed layer depth is difficult to determine.
Traditionally, there are two types of criteria for this
determination: the density difference criterion which will be used
in this work and the density gradient criterion. The first one
consists in estimating the mixed layer depth whose the
density is about the surface density increased by $0.01\hskip 0.1 cm
kg.m^{-3}$. This criterion is often used in the tropical area as in
Peters and al, 1988 \cite{Pe88}. So, the mixed layer is the one
where the density is inferior to the surface density to $0.01\hskip
0.1 cm kg.m^{-3}$. The second criterion is based on specifying
density gradient. The mixed layer depth is where the
density gradient is equal to the specified value.

\medskip
Mixing processes are intense in the homogeneous boundary layer, much
weaker near the pycnocline or in the deep ocean. The effects of
local fluxes of heat and momentum across the sea-surface induce
turbulence and then mixing processes in the surface layer.\ The
dynamical response of the ocean, and especially small scale
turbulence, has a significant effect on mixing processes.\ This is
true particularly in tropical areas. The rate of kinetic energy
dissipation and the vertical turbulent fluxes of heat and mass
cannot be measured directly, but they can be deduced from
measurements of vertical temperature gradients and horizontal
velocity \cite{Ph90}. Such experiments shown that turbulent
dissipation is higher near the equator than in low latitudes
\cite{Gr76}, \cite{Gr80}, \cite{Ga81}. Therefore in order to
modelize the mixed boundary layer and to represent correctly the
mixing processes, it is necessary to define a parametrization of
turbulent diffusion.\ The mixed layer being strongly dominated by
vertical fluxes, attention is drawn on vertical mixing which
requires a closure model in order to represent the Reynolds stress.
Recently, Goosse and al \cite{Go99} studied the sensitivity of a
global model to different parametrizations of vertical mixing.\ They
insist on the crucial role of mixing in the upper oceanic layers: it
has a direct impact on the sea surface temperature and then on ice
evolution, but affects also the vertical profile of velocity by
redistributing the momentum \cite{Bl93}.

\medskip
Classically, mixing parametrizations consist in the definition of
turbulent eddy viscosity and diffusivity coefficients. These
coefficients can be chosen either as constants or as fixed profiles
\cite{Hi96}. This is a simple but crude parametrization since
variations of mixing with time and location are forbidden.\ A more
suitable method is to define these coefficients as functions of
processes governing the mixing. In tropical oceans vertical
stratification and velocity shear are natural parameters since,
following Philander \cite{Ph90}, one of the reasons for higher
turbulent dissipation near the equator than in low latitudes, is the
large vertical shear observed in tropical currents. Pacanowski and
Philander \cite{Pa81} proposed a formulation for eddy viscosity and
diffusivity coefficients depending on the Richardson number which
represents the ratio between buoyancy effects and vertical shear.\
The Richardson number dependent formulations allow strong mixing in
high shear regions with low stratification, and low mixing
elsewhere.\ It has to be noted that stratification tends to reduce
turbulence and therefore mixing processes.

\medskip
In this paper, we study first order models. We focus 
on the behavior of Richardson number depending models: the basic one
proposed by Pacanowski-Philander (PP model) and two variants used
in Gent \cite{Gen91}, Blanke and Delecluse \cite{Bl93},Goosse and al
\cite{Go99}. In this study, the PP model is called R213 model and
the Gent model is called R23 model. The last model is named R224.
The studied region is the West Pacific Warm Pool. At
this location, the modelization based on the Richardson number is
realistic because of the large mean shear associated with the
equatorial undercurrent. Peters and al \cite{Pe88} have shown that
the PP model underestimates the turbulent mixing at low Richardson
number, while overestimating the turbulent mixing at high Richardson
number in comparison with turbulent measurements.  They found that
the simulated thermocline is much too diffused compared with
observations. Furthermore, this
scheme overestimates the surface current intensity. Also, the simulated
equatorial undercurrent is too shallow compared to observations
(\cite{Ha95},\cite{Ni95},\cite{Li01}). Halpern and al \cite{Ha95}
compared the PP scheme and the MY scheme. At the equator, the
current and temperature simulated by the PP scheme are more
realistic than those simulated by the Mellor
and Yamada scheme. The Gent model, one of three studied ones, gives
realistic results in the West Pacific Warm Pool. This model simulates a
sharp thermocline which is in agreement with the observations
\cite{Gen91}. Furthermore, it gives good results for the annual average
SST (Sea Surface Temperature) at the equator.

\medskip
In this paper, we investigate in case of each model, the simulated
mixed layer depth, the form of the simulated pycnocline (sharp or
diffuse) and the intensity of the simulated surface current. The
numerical implementation is based on a non-conservative numerical
scheme. We initialize
the code with realistic velocity and density profiles. These
profiles come from to the TOGA-TAO array (McPhaden, 1995,
\cite{Mc95}). The geographic location is $0^{\circ}N,165^{\circ}E$
which is found in the West Pacific Warm Pool. In the first part, we study a mixed 
layer induced by the wind stress.
Then, we initialize the code
with a density profile showing a static instability zone. At last, we simulate a long time case.

\section{Modelization of the mixed layer}

\subsection{Setting of the problem}

Variables used in this paper to describe the mixed layer are
statistical means of the horizontal velocity and of the density
denoted by $ \left( u,v,\rho \right) .$ In the ocean, the density is
a function of temperature and the salinity through a state equation.
So, we consider the density as an idealized thermodynamic variable
which is intended to represent temperature and salinity variations.\
The formation \ of the mixing layer is a response to sea-air
interactions: wind-stress, solar heating, precipitations or
evaporation acting on the sea surface.\ The variability of
temperature is considered as essential to understand the response of
the ocean.\ For example, the existence of a sharp pycnocline is a
well known feature of the tropical areas.\ The thermal inertia of
the water column is linked with the depth of the pycnocline, which
influences the sea-surface temperature. The role of the haline
stratification, sometimes considered as less important, has been
recently evidenced by Vialard and Delecluse \cite{Vi98}: a numerical
modelization of the tropical Pacific produces a ''barrier layer''
depending on surface forcings and large scale circulation.\ The term
''barrier layer'' refers to the water column located between the
bottom of the mixed layer and the top of the pycnocline. It is
present when the isohaline layer is shallower than the isothermal
layer and then the depth of the mixed layer is controlled by the
salinity.

The model studied hereafter is not expected to describe all the
phenomena occuring in the mixed layer.\ Its purpose is only a better
understanding of a classical closure model.\ Therefore we shall use
simplified equations governing the variables $u$ (zonal
velocity),$v$ (meridian velocity) and $\rho $ (density).

The mixed layer being strongly dominated by vertical fluxes, we
shall suppose that $u$,$v$ and $\rho $ are horizontally homogeneous
and we so obtain a one-dimensional modelization.\ The Coriolis force
will be neglected, which is valid in tropical oceans. Therefore, the
equations governing the mixing layer are
\begin{equation}
\left\{
\begin{array}{c}
\dfrac{\partial u}{\partial t}=-\dfrac{\partial }{\partial
z}\left\langle
u^{\prime }\,w^{\prime }\right\rangle ,\medskip  \\
\dfrac{\partial v}{\partial t}=-\dfrac{\partial }{\partial
z}\left\langle
v^{\prime }\,w^{\prime }\right\rangle ,\medskip  \\
\dfrac{\partial \rho }{\partial t}=-\dfrac{\partial }{\partial z}%
\left\langle \rho ^{\prime }\,w^{\prime }\right\rangle ,
\end{array}
\right.
\end{equation}
where $u^{\prime },v^{\prime },w^{\prime },\rho ^{\prime }$
represent the fluctuations
of the horizontal velocity, vertical velocity and the density.\ The notation $%
\left\langle \hspace{0.3cm}\right\rangle $ signifies that the
quantity is statistically averaged.\ Equations (1) are the classical
equations corresponding to a modelization of a water column. Such
equations can be found in \cite{Les97}
 as the equations of the boundary layer.
Equations (1) are not closed and then the vertical fluxes appearing
in the right-hand side have to be modelized.\

\medskip
We study in this paper the behavior of a very classical closure
modelization that uses the concept of eddy coefficients in order to
represent turbulent fluxes.\ So we set
\begin{eqnarray*}
-\left\langle u^{\prime }\,w^{\prime }\right\rangle  &=&\nu _{1}\dfrac{%
\partial u}{\partial z}, \\
-\left\langle v^{\prime }\,w^{\prime }\right\rangle  &=&\nu _{1}\dfrac{%
\partial v}{\partial z}, \\
-\left\langle \rho ^{\prime }\,w^{\prime }\right\rangle  &=&\nu _{2}\dfrac{%
\partial \rho }{\partial z}.
\end{eqnarray*}
Coefficients $\nu _{1}$ and $\nu _{2}$ are called vertical eddy
viscosity and diffusivity coefficients and will be expressed as
functions of the Richardson number $R$ defined as
\begin{equation*}
{\displaystyle R=-}\dfrac{g}{\rho_{0}}{\displaystyle \,}\dfrac{\tfrac{\partial \rho }{%
\partial z}}{\left( \tfrac{\partial u}{\partial z}\right) ^{2}+\left( \tfrac{\partial v}{\partial z}\right) ^{2}}{\LARGE ,}
\end{equation*}
where $g$ is the gravitational acceleration and $\rho_{0}$ a
reference density (for example $\rho_{0}=1025\hskip 2 pt
kg.m^{-3})$.

The set of equations, initial and boundary conditions governing the
mixing layer can now be written

\medskip

\begin{equation}\label{syst}
\left\{
\begin{array}{l}

\medskip
\dfrac{\partial u}{\partial t}-\dfrac{\partial }{\partial z}\left( \nu _{1}%
\dfrac{\partial u}{\partial z}\right) =0,\\

\medskip
\dfrac{\partial v}{\partial t}-\dfrac{\partial }{\partial z}\left( \nu _{1}%
\dfrac{\partial v}{\partial z}\right) =0,\\

\medskip
\dfrac{\partial \rho }{\partial t}-\dfrac{\partial }{\partial
z}\left( \nu
_{2}\dfrac{\partial \rho }{\partial z}\right) =0,\text{ for }t\geqslant 0%
\text{ and }-h\leqslant z\leqslant 0,\\

\medskip
u=u_{b},\,\,v=v_{b},\,\,\rho =\rho _{b}\text{ \ at the depth }z=-h, \\

\medskip
\displaystyle \nu _{1}\dfrac{\partial u}{\partial z}=\frac{\rho_{a}}{\rho_{0}}V_{x},\,\,\nu _{1}\dfrac{\partial v}{\partial z}=\frac{\rho_{a}}{\rho_{0}}V_{y},\,\, \nu _{2}\dfrac{\partial \rho }{%
\partial z}=Q\text{ \ at the surface }z=0, \\

\medskip
u=u_{0},\,\,v=v_{0},\,\,\rho =\rho _{0}\text{ \ at initial time
}t=0.
\end{array}
\right.
\end{equation}

\medskip

In system (2), the constant $h$ denotes the thickness of the studied
layer that must contain the mixing layer.\ Therefore the circulation
for $z<-h$, under the boundary layer, is supposed to be known,
either by observations or by a deep circulation numerical model.
This justifies the choice of Dirichlet boundary conditions at
$z=-h$, $u_{b}$, $v_{b}$ and $\rho _{b}$ being the values of
horizontal velocity and density in the layer located below the mixed
layer.\ The air-sea interactions are represented by the fluxes at
the sea-surface: $V_{x}$ and $V_{y}$ are respectively the forcing
exerced by the zonal wind-stress and the meridional wind-stress and
$Q$ represents the thermodynamical fluxes, heating or cooling,
precipitations or evaporation. We have $V_{x}=C_{D}\left|
u^{a}\right| ^{2}$ and $V_{y}=C_{D}\left| v^{a}\right| ^{2}$, where
$U^{a}=(u_{a},v_{a})$ is the air velocity and $C_{D}(=1,2.10^{-3})$
a friction coefficient.

\medskip
We study hereafter three different formulations for the eddy coefficients $%
\nu _{1}=f_{1}\left( R\right) $ and $\nu _{2}=f_{2}(R).$ Functions
$f_{1}$ and $f_{2}$ can be defined as
\begin{equation}
f_{1}\left( R\right) =\alpha _{1}+\dfrac{\beta _{1}}{\left( 1+5R\right) ^{2}}%
,\,\,\,\,f_{2}\left( R\right) =\alpha _{2}+\dfrac{f_{1}\left( R\right) }{1+5R%
}=\alpha _{2}\text{+}\frac{\alpha _{1}}{1+5R}\text{+}\frac{\beta _{1}}{%
\left( 1+5R\right) ^{3}}.
\end{equation}
Formulation (3) corresponds to the modelization of the vertical
mixing proposed by Pacanowski and Philander \cite{Pa81}. They
proposed for coefficients $\alpha _{1},\beta _{1}$and $\alpha _{2}$
the following values: $\ \ \ \alpha _{1}=1.10^{-4},\,\,\beta
_{1}=1.10^{-2},\,\,\ \alpha _{2}=1.10^{-5}($units:\hskip 0.1
cm$m^{2}s^{-1})$. This formulation has been used in the OPA code
developed by Paris 6 University \cite{Bl93},\cite{Ma97}
 with coefficients $\alpha _{1}=1.10^{-6},\,\,\beta
_{1}=1.10^{-2},\,\,\alpha _{2}=1.10^{-7}\left( \text{units:\thinspace }%
m^{2}s^{-1}\right) $.The selection criterion for the coefficients
appearing in these formulas was the best agreement of numerical
results with observations carried out in different tropical areas.

A variant of formulation (3),proposed by Gent \cite{Gen91}, is
$\newline
$%
\begin{equation}
f_{1}\left( R\right) =\alpha _{1}+\dfrac{\beta _{1}}{\left( 1+10R\right) ^{2}%
},\,\,\,\,f_{2}\left( R\right) =\alpha _{2}+\dfrac{\beta
_{2}}{\left( 1+10R\right) ^{3}}
\end{equation}
with $\alpha _{1}=1.10^{-4},\,\,\beta _{1}=1.10^{-1},\,\,\ \alpha
_{2}=1.10^{-5},\,\,\beta _{2}=1.10^{-1}($units: $m^{2}s^{-1}).$ A
formulation similar to $\left( 4\right) \,$when replacing $10R$ by
$5R$ and varying the values of the coefficients $\alpha _{1},\alpha
_{2\text{ }}$between the surface and the depth 50m is used in
\cite{Go99}.

In this paper, we will also study the properties of another
formulation close to formula (3):
\begin{equation}
f_{1}\left( R\right) =\alpha _{1}+\dfrac{\beta _{1}}{\left( 1+5R\right) ^{2}}%
,\,\,\,\,f_{2}\left( R\right) =\alpha _{2}+\dfrac{f_{1}\left( R\right) }{%
(1+5R)^{2}}=\alpha _{2}+\frac{\alpha _{1}}{(1+5R)^{2}}+\frac{\beta _{1}}{%
\left( 1+5R\right) ^{4}},
\end{equation}

with $\alpha _{1}=1.10^{-4},\,\,\beta _{1}=1.10^{-2},\,\,\alpha
_{2}=1.10^{-5},\,\ \beta _{2}=1.10^{-3}\left( \text{units: \thinspace }%
m^{2}s^{-1}\right).$

Eddy viscosity $\nu _{1\text{ }}$defined by (5) is the same
as the coefficient given by Pacanowski and Philander. The definition
of the eddy
diffusivity coefficient $\nu _{2}$ differs by the exponent of the term $%
(1+5R)$. 

\medskip
In formulas (3) to (5), the eddy coefficients $\nu _{1}$ and $\nu
_{2}$ are defined as functions of the Richardson number R through
the terms $(1+\gamma R)^{n}$ appearing at the denominator.
Hereafter, these three formulations will be denoted respectively by
R213, R23 and R224 where R signifies Richardson number and the
integer values are the exponents of $(1+\gamma R)$ in the
definitions of $\nu _{1}$ and $\nu _{2}.$

Eddy coefficients defined by relations (3) to (5) all present a
singularity for a negative value of the Richardson number $R=-0.2$
or $R=-0.1.$ We have
plotted in Figure 1a the curves $\nu _{1}=f_{1}(R)$. In formulations $%
\left( 3\right) $ or $\left( 4\right) $ the coefficient of eddy diffusivity $%
\nu _{2}$ becomes negative for values of $R$ lower than $-0.2$ or
$-0.1$, and therefore the models are no more valid. The curves $\nu
_{2}=f_{2}\left( R\right) $ obtained with formulations (3) (4) (5) and
are plotted in Figure 1b.

\newpage

\begin{center}

\includegraphics[scale=0.45]{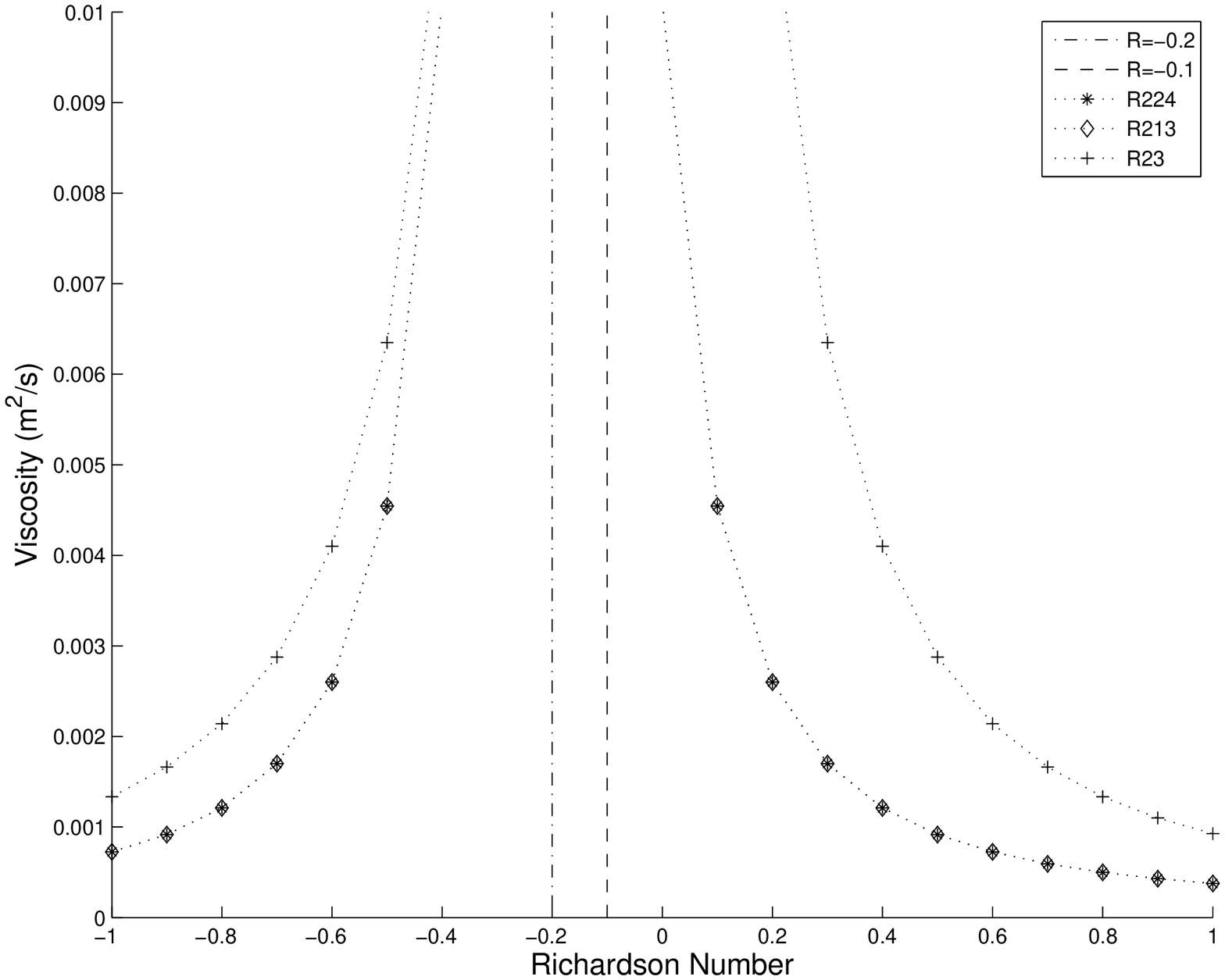}

\textbf{Figure 1a}:
Viscosity ($\nu_1=f_1(R)$) for all models\\

\end{center}

\begin{tabular}{ccc}
\hskip -2.5cm \includegraphics[scale=0.42]{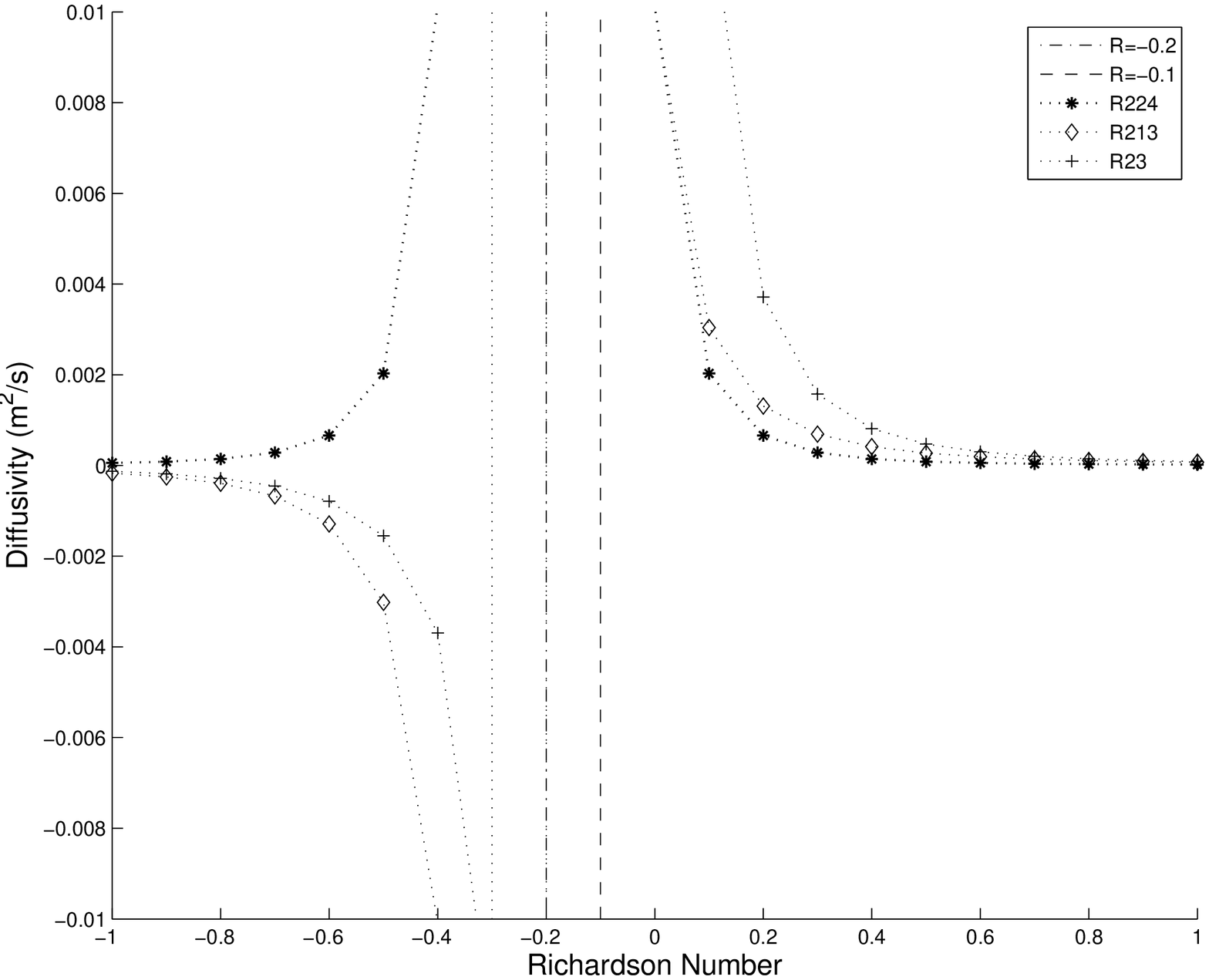}&&
\includegraphics[scale=0.42]{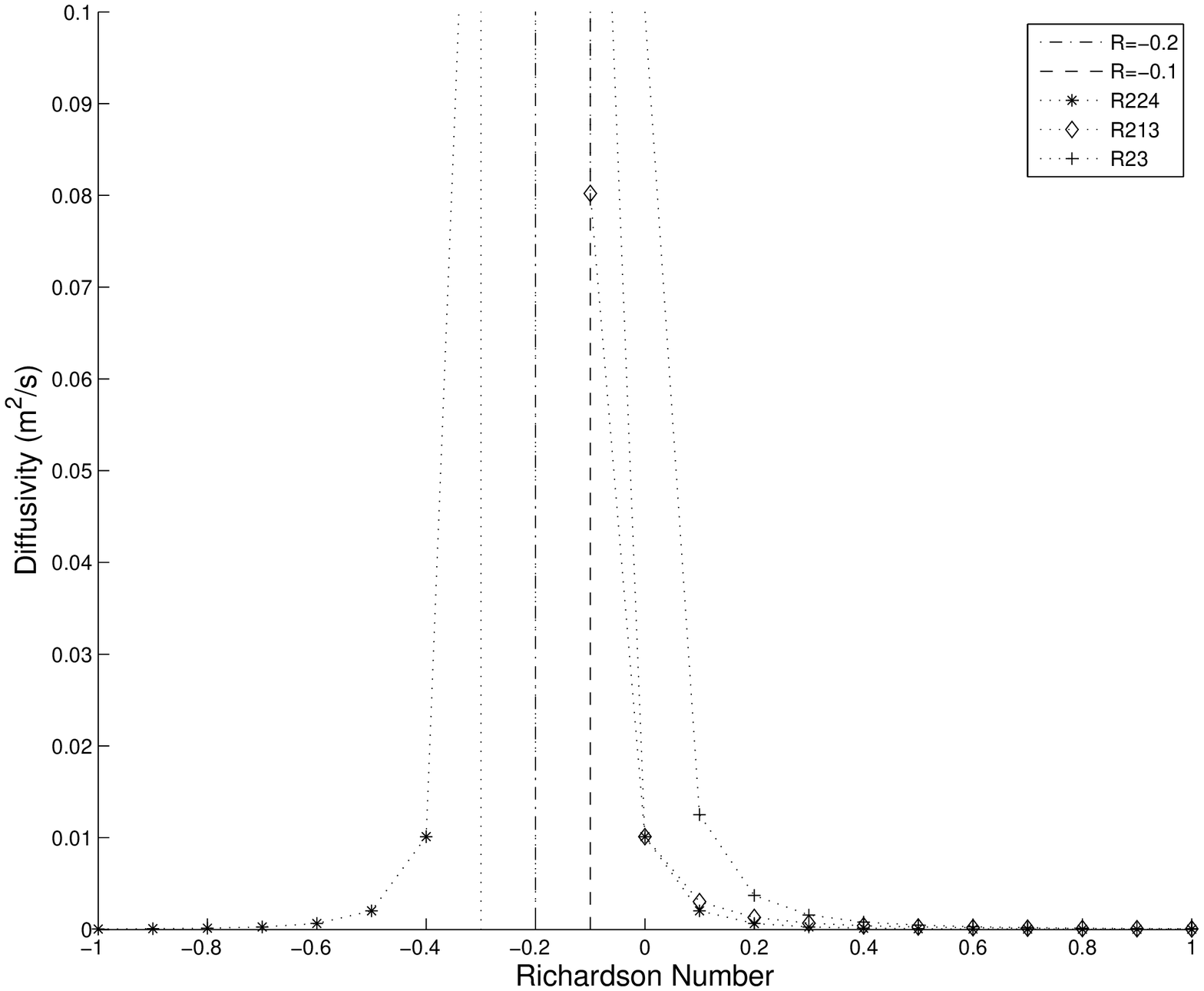}
\end{tabular}

\begin{center}
\textbf{Figure 1b}:
Diffusivity ($\nu_2=f_2(R)$) for all models\\
\end{center}

Problem $\left( 2\right) $ coupled with one of the definitions (3)
to (5) for eddy coefficients retains the vertical shear and buoyancy
effect which are two important processes for the generation of the
mixed boundary layer, especially in tropical areas.


\newpage

\section{Numerical Experiments}

\subsection{Finite Difference Scheme}

We want to resolve numerically the system (\ref{syst}). We replace
the continuous variables ($u,v,\rho,\nu_{1},\nu_{2}$) by discrete
variables
($u_{i}^{n},v_{i}^{n},\rho_{i}^{n},\left(\nu_{1}\right)_{i}^{n},\left(\nu_{2}\right)_{i}^{n}$)
which are the approximate solutions at time $n\Delta t$ (with
$n=1,2,...,N$) and at points $(i-NI)\Delta z$ (with $i=1,2,...,NI$).
We discretize the 1D-domain in z-levels where z is the vertical
coordinate.

\medskip
We use a second-order central difference scheme for the second space
derivative and a first-order backward difference scheme for the
first space derivative. These previous schemes can be written as:

\begin{itemize}
\item Second-order central difference scheme:
$\displaystyle \left(\frac{\partial^{2}{u}}{\partial{z^{2}}}\right)_{i}^{n+1}=\frac{u_{i+1}^{n+1}-2u_{i}^{n+1}+u_{i-1}^{n+1}}{\Delta z^2}$
\item First-order backward difference scheme:
$\displaystyle \left(\frac{\partial{u}}{\partial{z}}\right)_{i}^{n+1}=\frac{u_{i}^{n+1}-u_{i-1}^{n+1}}{\Delta z}$

\end{itemize}

The grid spacing, $\Delta z$, is equal to $5 \hskip 0.1 cm m$ in cases 2 and 3 and equal to $1 \hskip 0.1 cm m$
in case 1.
The time step, $\Delta t$, is equal to $60 \hskip 0.1 cm s$. In
time, we use implicit velocities and implicit density. The
viscosity ($\nu_{1}$) and diffusivity ($\nu_{2}$) are explicit. The
basin depth is $100 \hskip 0.1 cm m$. The boundary conditions are
treated with a first-order backward difference scheme. We choose
Neumann boundary conditions at the surface and Dirichlet boundary
conditions in $z=-h$.

\medskip
The numerical scheme is the following:

\begin{equation}\label{sc}
\left\{
\begin{array}{l}
\displaystyle\frac{u_{i}^{n+1}-u_{i}^{n}}{\Delta t}-\left(\frac{\nu_{1})_{i}^{n}-\nu_{1})_{i-1}^{n}}{\Delta z}\right)\cdot \left(\frac{u_{i}^{n+1}-u_{i-1}^{n+1}}{\Delta z}\right)-\nu_{1})_{i}^{n}\cdot\left(\frac{u_{i+1}^{n+1}-2u_{i}^{n+1}+u_{i-1}^{n+1}}{\Delta z^2}\right)=0, \\\\

\displaystyle\frac{v_{i}^{n+1}-v_{i}^{n}}{\Delta t}-\left(\frac{\nu_{1})_{i}^{n}-\nu_{1})_{i-1}^{n}}{\Delta z}\right)\cdot \left(\frac{v_{i}^{n+1}-v_{i-1}^{n+1}}{\Delta z}\right)-\nu_{1})_{i}^{n}\cdot\left(\frac{v_{i+1}^{n+1}-2v_{i}^{n+1}+v_{i-1}^{n+1}}{\Delta z^2}\right)=0, \\\\

\displaystyle\frac{\rho_{i}^{n+1}-\rho_{i}^{n}}{\Delta t}-\left(\frac{\nu_{2})_{i}^{n}-\nu_{2})_{i-1}^{n}}{\Delta z}\right)\cdot \left(\frac{\rho_{i}^{n+1}-\rho_{i-1}^{n+1}}{\Delta z}\right)-\nu_{2})_{i}^{n}\cdot\left(\frac{\rho_{i+1}^{n+1}-2\rho_{i}^{n+1}+\rho_{i-1}^{n+1}}{\Delta z^2}\right)=0.

\end{array}
\right.
\end{equation}

and the Neumann boundary conditions at the surface are computed as: \\

 $\displaystyle\nu_{1})_{i}^{n}\cdot\left(\frac{u_{i}^{n+1}-u_{i-1}^{n+1}}{\Delta z}\right)=\frac{\rho_{a}}{\rho_{0}}V_{x},  \hskip 0.5cm
 \displaystyle\nu_{1})_{i}^{n}\cdot\left(\frac{v_{i}^{n+1}-v_{i-1}^{n+1}}{\Delta z}\right)=\frac{\rho_{a}}{\rho_{0}}V_{y}, \hskip 0.5cm
\displaystyle\nu_{2})_{i}^{n}\cdot\left(\frac{\rho_{i}^{n+1}-\rho_{i-1}^{n+1}}{\Delta z}\right)=Q.         $\\

Moreover, the residual values are computed as:

\begin{equation}\label{res}
\begin{array}{l}
\displaystyle  r^{n}=\left(\sum_{i=1}^{NI} \left|u_{i}^{n+1}-u_{i}^{n}\right| ^{2}+\sum_{i=1}^{NI} \left|v_{i}^{n+1}-v_{i}^{n}\right| ^{2}+ \sum_{i=1}^{NI} \left|\rho_{i}^{n+1}-\rho_{i}^{n}\right| ^{2} \right)^{1/2}. \\
\end{array}
\end{equation}

Our numerical scheme is non-conservative but it produces the same results as those given by a conservative one. Furthermore, the residual values decrease monotonically to zero as the flow reaches the steady states, better than the conservative scheme where the behaviour of the residual is more erratic.

\subsection{Numerical Results}

In this section, we aim to study an Equatorial Pacific region called
the West-Pacific Warm Pool. So, we initialize the code with data
from the TAO array (McPhaden \cite{Mc95}). It is located at the
equator between $120^{\circ}E$ and $180^{\circ}E$. The sea
temperature is high and almost constant along the year ($28-30^\circ
C$). The precipitation are intense and hence the salinity is low. In the first part, we study a mixed 
layer induced by the wind stress. Then, we consider a case where the code is initialized 
by a density profile showing a static instability zone. Finally, we simulate a long time case.
All our numerical results are grid-size ($\Delta z$) and time step ($\Delta t$) independent, in the sense that they remain practically unchanged as $\Delta z$ and $\Delta t$ decrease.

\subsubsection{\label{init}The initial data}

We use data available from the Tropical Atmosphere Ocean (TAO) array
(McPhaden \cite{Mc95}). The TAO project aims to study the exchange
between the tropical oceans and the atmosphere. The TAO data have
being very used in numerical simulations. The velocity data comes
from the ACDP (Acoustic Doppler Current Profiler) measurements. To
obtain the appropriate profiles, we interpolate the data by a
one-order linear interpolation. We initialize our code by these
profiles at $0^\circ N, 165^\circ E$ and we obtain the results below.
The buoyancy flux is equal to $-1.10^{-6} \hskip 0.1 cm
kg.m^{-2}.s^{-1}$ ($\simeq -11 \hskip 0.1 cm W/m^{2}$) in all cases. This heat flux is in agreement
with Gent \cite{Gen91} because the heat flux must be in the
range [$0 \hskip 0.1 cm W/m^{2}$ ; $20 \hskip 0.1 cm W/m^{2}$]  between
$140^{\circ}E-180^{\circ}E$ and $10^{\circ}N-10^{\circ}S$. 

%


\subsubsection{Case 1: A mixed layer induced by the wind stress}

We use, in this case, as initial data, the TAO's data for the
time period between the 15th June 1991 and the 15th July 1991. The initial
profiles are displayed on figure 2. The initial zonal velocity
profile presents a westward current at the surface and, below it, an
eastward undercurrent whose maximum is located about $55 \hskip 0.1
cm m$. Deepest, we observe a westward undercurrent. The initial
density profile does not display a mixed layer.

\hskip -2.3cm\includegraphics[scale=0.425]{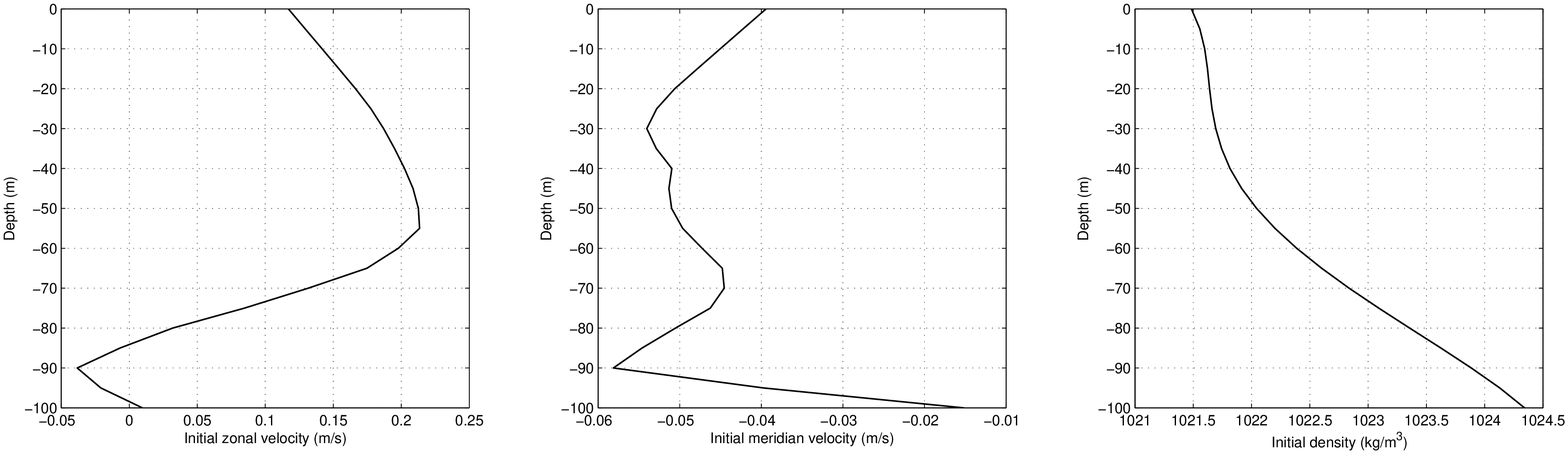}

\begin{center}
\textbf{Figure 2}: Initial zonal velocity, meridian velocity and density profiles (from left to right). 
\end{center}

The buoyancy flux is equal to $-1.10^{-6} \hskip 0.2 cm
kg.m^{-2}.s^{-1}$ ($\simeq -11W/m^2$). The model is integrated for 48 hours. 
The grid spacing is equal to 1 meter and the time step is equal to 60 seconds.
The chosen wind stress is stronger than in reality because we want to simulate 
a mixed layer induced by the wind stress.
These values correspond to an another period in the studied
year. The zonal wind is equal to $8.1 \hskip 0.1 cm m/s$ (eastward
wind) and the meridional wind is equal to $2.1 \hskip 0.1 cm m/s$
(northward wind).

\medskip
The numerical results are displayed on figure 4. 
The final density profile displays a similar mixed layer
for R213, R224, R23 models. Furthermore, the pycnocline simulated by R224,
R213 and R23 models are similar.

\medskip
The deep flow ($60-100 \hskip 0.1
cm m$) are the same for all models because the surface fluxes are not
strong enough to affect the deep water.

\medskip
The R213 and R224 surface current are of the same kind. The R23 model underestimates this current.
We notice an increase in zonal and meridian surface current in comparison with the initial profiles
which concord with a south-westerly wind.
The surface current behavior can be explained by the viscosity and
diffusivity values. The final
diffusivity and viscosity are displayed on figure 3 for all models.
We observe the order described below.

$$\left(\nu_1\right)_{23}>\left(\nu_1\right)_{213}>\left(\nu_1\right)_{224}$$
$$\left(\nu_2\right)_{23}>\left(\nu_2\right)_{213}>\left(\nu_2\right)_{224}$$

The R23 model has a strongest viscosity and diffusivity. Therefore,
the R23 surface current is lower than the other surface currents. 
As the R224 and R213 viscosities and R224 and R213 diffusivities are of the same kind,
the R224 and R213 surface currents are similar. However, the R224
surface current is slightly stronger than the R213 surface current.

\begin{tabular}{ccc}
\includegraphics[scale=0.35]{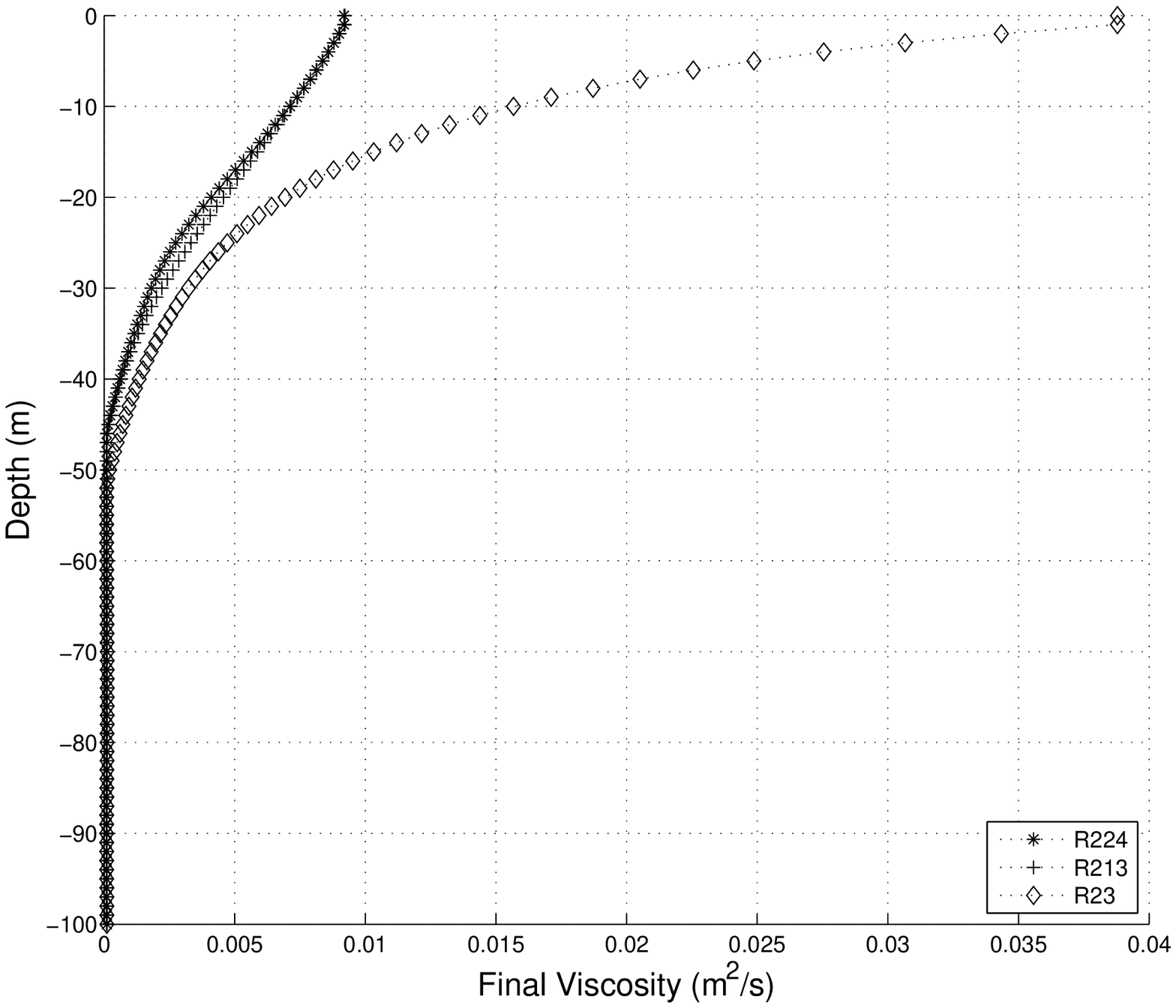} & &
\includegraphics[scale=0.35]{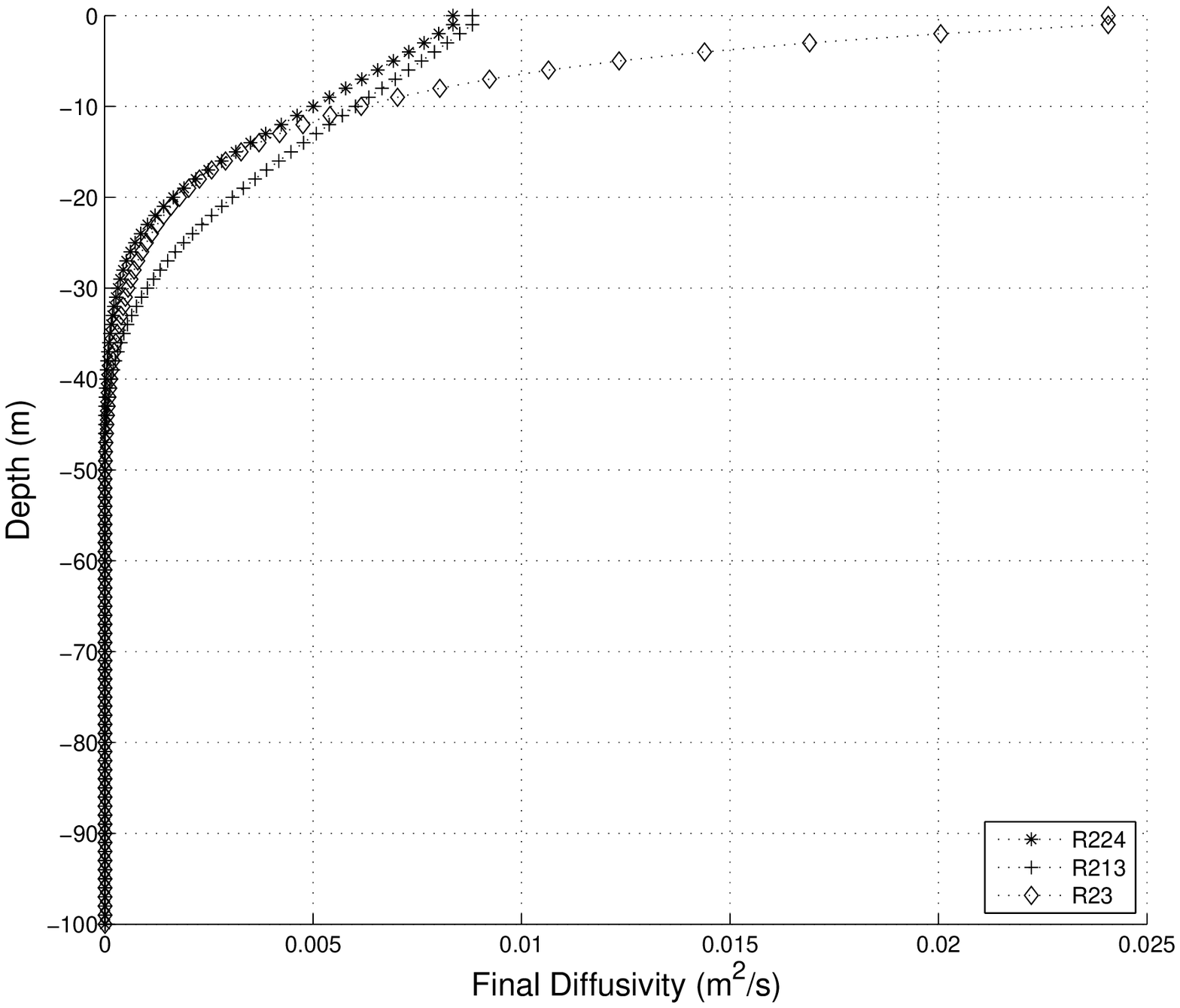}
\end{tabular}

\begin{center}
\textbf{Figure 3}: Final viscosity (left-hand side) and final
diffusivity (right-hand side) for all models.
\end{center}

\hskip -2.3cm \includegraphics[scale=0.4]{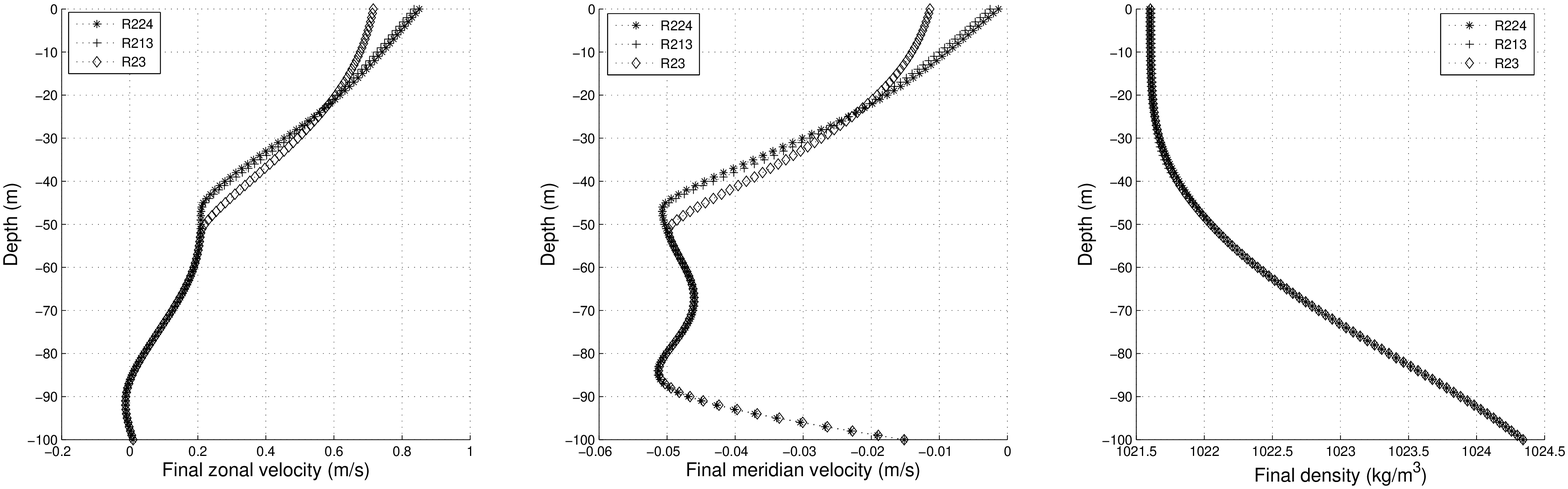}
\medskip
\hskip -2.3cm \includegraphics[scale=0.4]{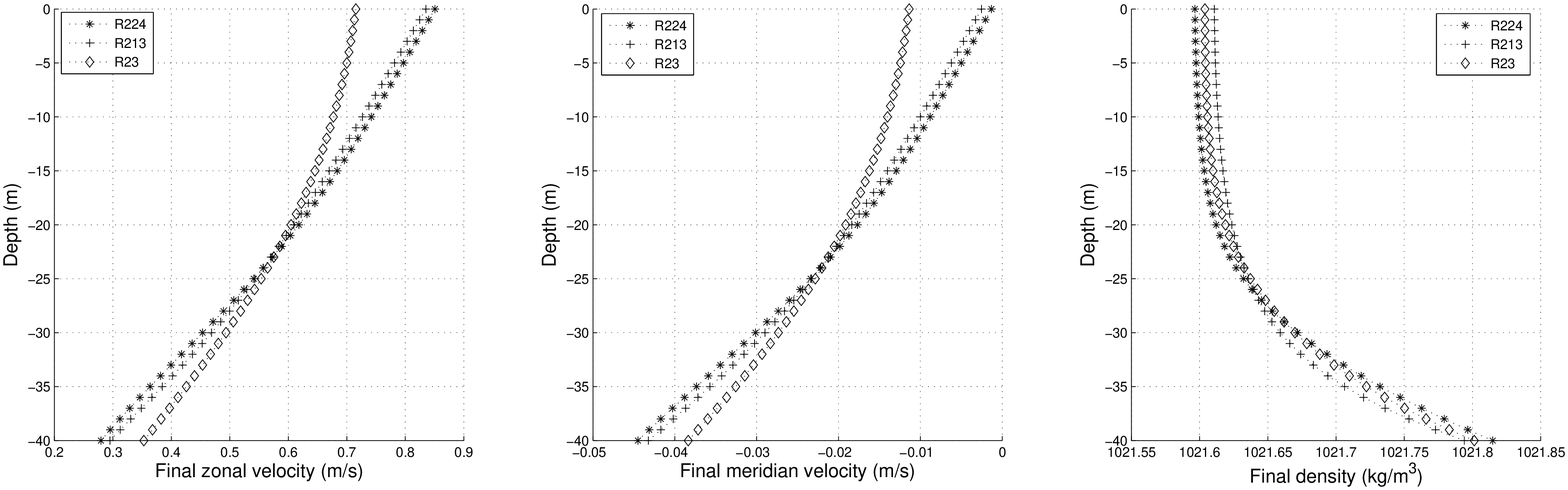}

\begin{center}
\textbf{Figure 4}: Comparison of three turbulence models. We can see, on the top row, the entire water column and on the bottom row, the shallow flow.
We have plotted, from left to right, the finals zonal velocity profiles, the final meridian velocity profiles and the final density profiles.
\end{center}

\medskip

\begin{center}

\includegraphics[scale=0.48]{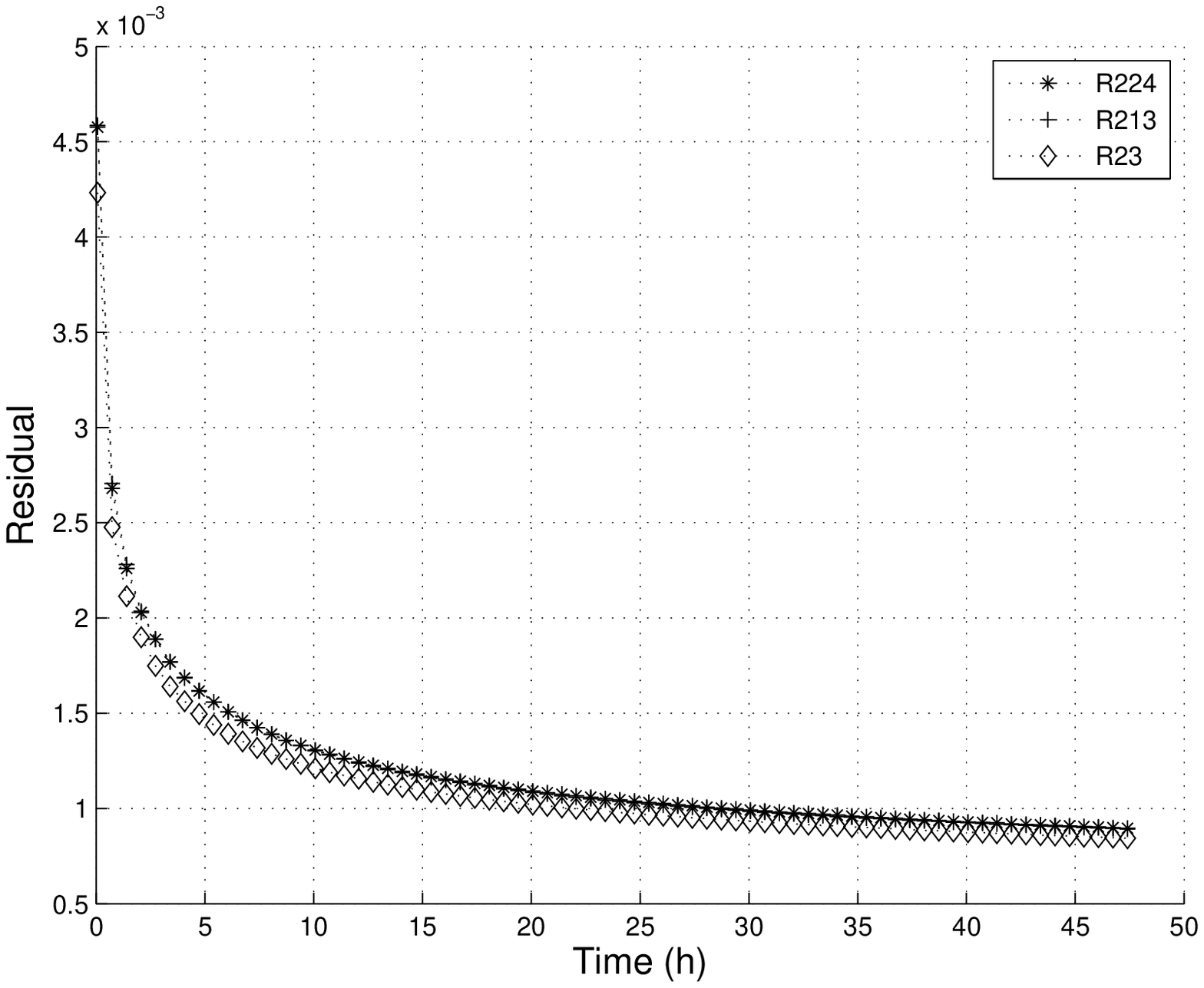}

\textbf{Figure 5}: Comparison of residual values.
\end{center}

On figure 5, we notice a monotonic numerical convergence to the steady state for all models.


\subsubsection{Case 2: A static instability zone in the initial density profile}

The code is initialized with the 17th November 1991 data. The initial
zonal velocity profile displays an eastward current whose maximum is
located about $55 \hskip 2 pt m$. The initial meridian velocity
profile displays a southward current whose maximum is located about
$20 \hskip 2 pt m$. The initial density profile displays a static
instability zone between $-30 \hskip 0.1cm m$ and $-50 \hskip 0.1cm
m$. Notice that there is a similar instability zone for the following days: 16th November,
the 19th November, the 20th November, the 21th November and the 22th November 1991.
Here, we observe a seventy meters deep mixed layer. However,
this mixed layer is not homogeneous. The initial profiles are
displayed on figure 6.

\medskip
\hskip -2.5cm\includegraphics[scale=0.36]{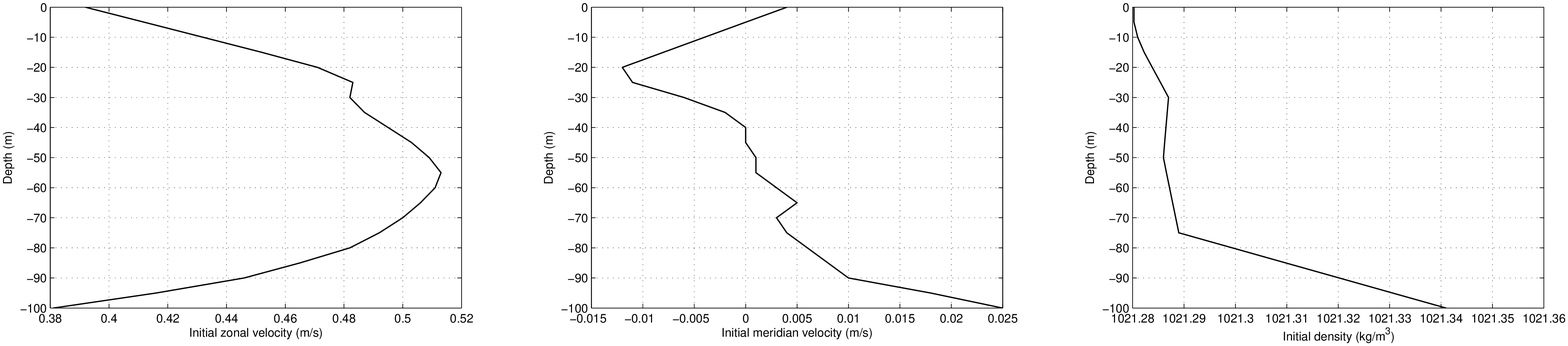}

\begin{center}
\textbf{Figure 6}:
The initial profiles for zonal velocity, meridian velocity and density (from left to right). \\
\end{center}

On figure 7, the initial Richardson number, called $R^{0}$, is near
to $-0.2$ for $z=-45 \hskip 0.1 cm m$. The R213 and R224 models
have infinite viscosity and diffusivity for $R^{0}=-0.2$ (see Figures 1a and 1b).
Hence, the
R213 diffusivity (see figure 9) and the R224 diffusivity (see
figure 8) are large for this depth. The initial Richardson number
(see figure 7) is inferior to $-0.2$ for $z=-35 \hskip 0.1 cm m$
and $z=-50 \hskip 0.1 cm m$. Therefore, the R213 diffusivity is
negative for this depths (see Figure 1b). In the range $[-35\hskip 0.1cm m,-50 \hskip 0.1 cm m]$,
the initial richardson number is inferior to $-0.1$ and hence the
R23 diffusivity is negative (see Figure 1b). On figure 9, the negative diffusivities
are marked by a point for R23 and R213 models. The
R224 diffusivity (see figure 8) is always positive. In
physical reality, negative diffusivity does not exist. The
diffusivity was estimated by Osborn and Cox \cite{Os72} with
measurements of very small scale vertical structure. They have shown  that in our studied
region, the diffusivity was in the range [$1.10^{-2} \hskip 0.1 cm
cm^{2}.s^{-1}$, $1.10^{3} \hskip 0.1 cm cm^{2}.s^{-1}$]. 
So, we can not use R213 and R23
models in this case.

\begin{tabular}{ccc}
\hskip -1cm\includegraphics[scale=0.44]{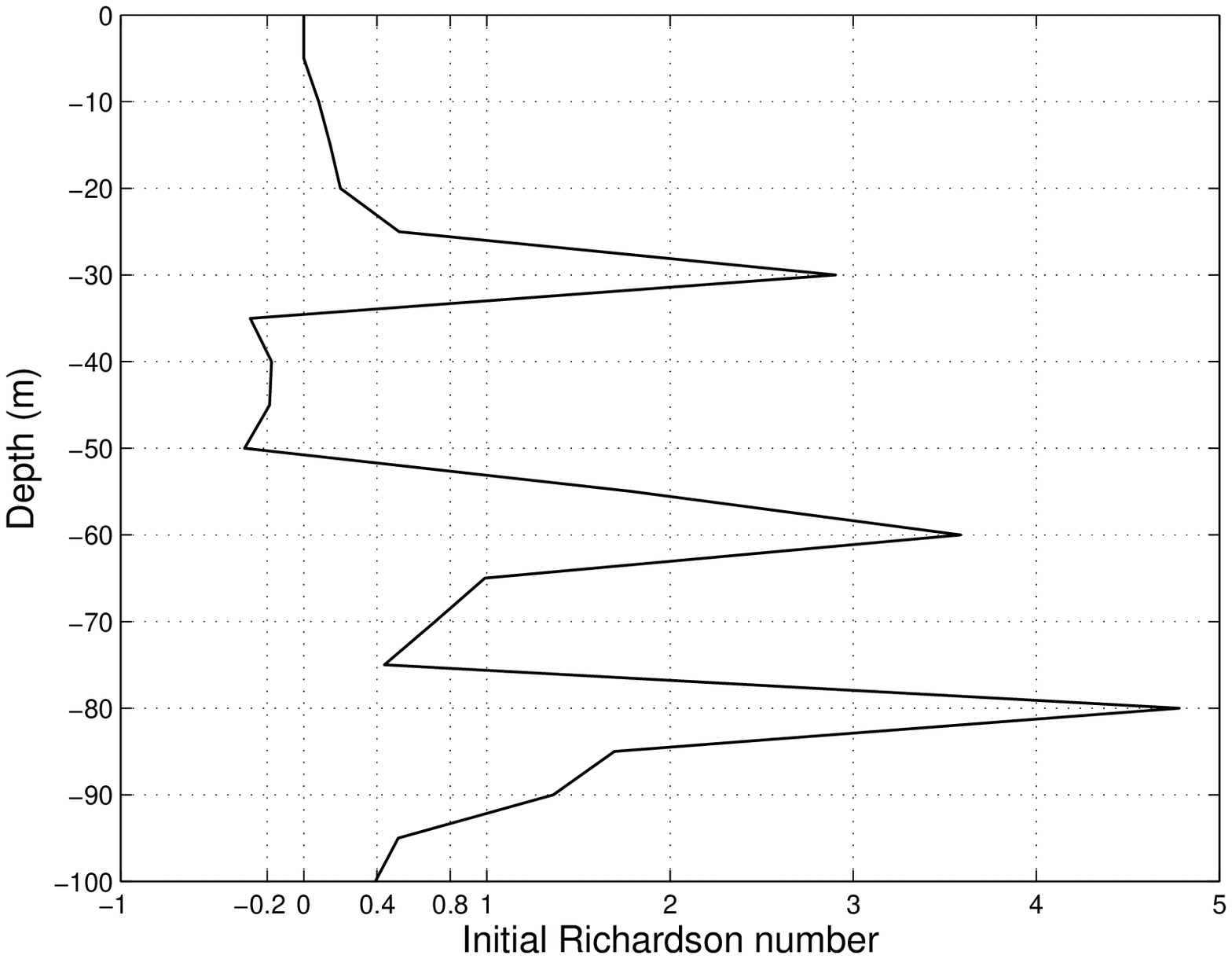} & &
\includegraphics[scale=0.44]{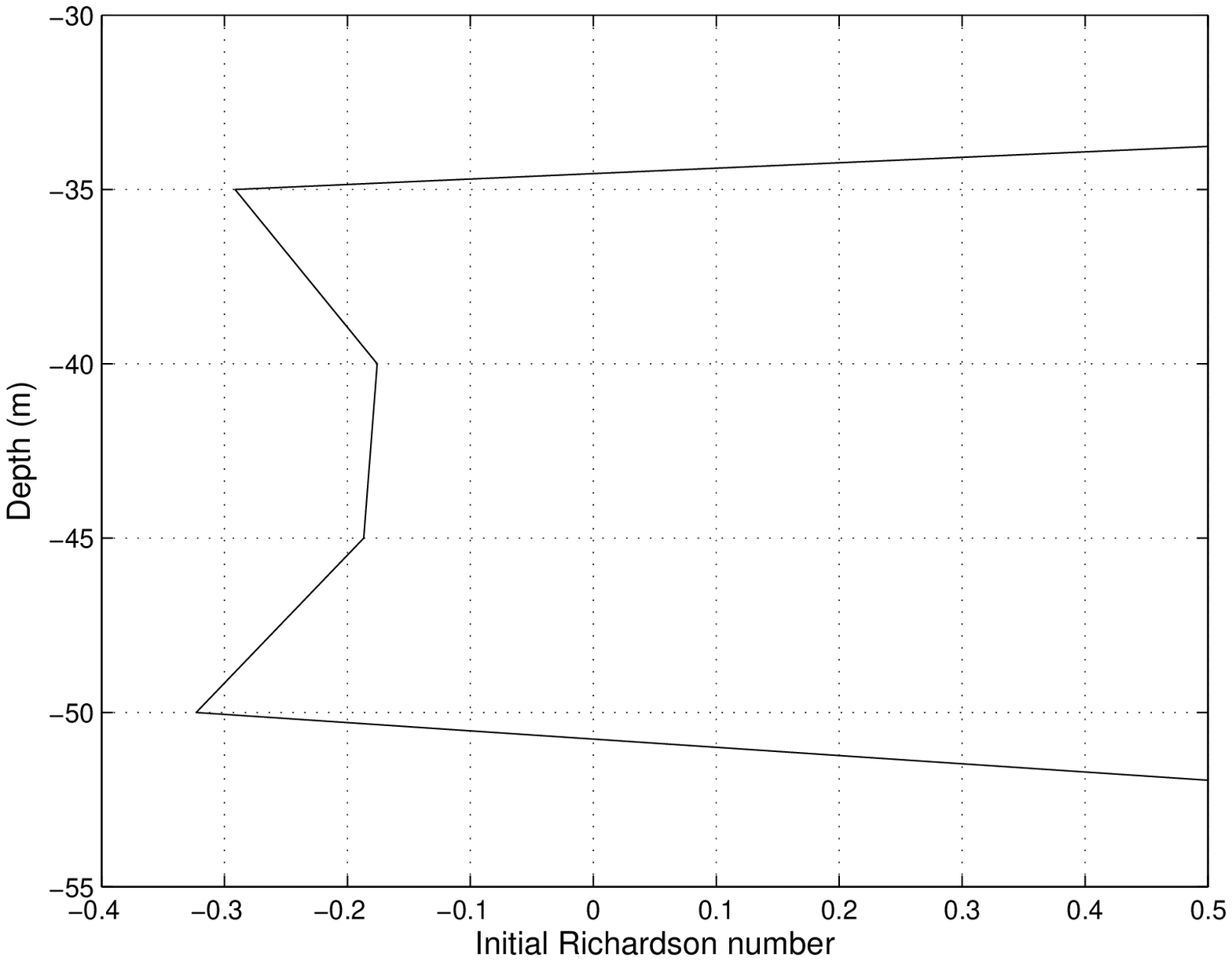}
\end{tabular}

\begin{center}
\textbf{Figure 7}: The initial richardson number for different
depths. We can see a entire water column on the left-hand side and the flow between -30m and -55m on the 
right-hand side.
\end{center}

\begin{center}
\begin{tabular}{ccc}
\hskip -1.2cm\includegraphics[scale=0.44]{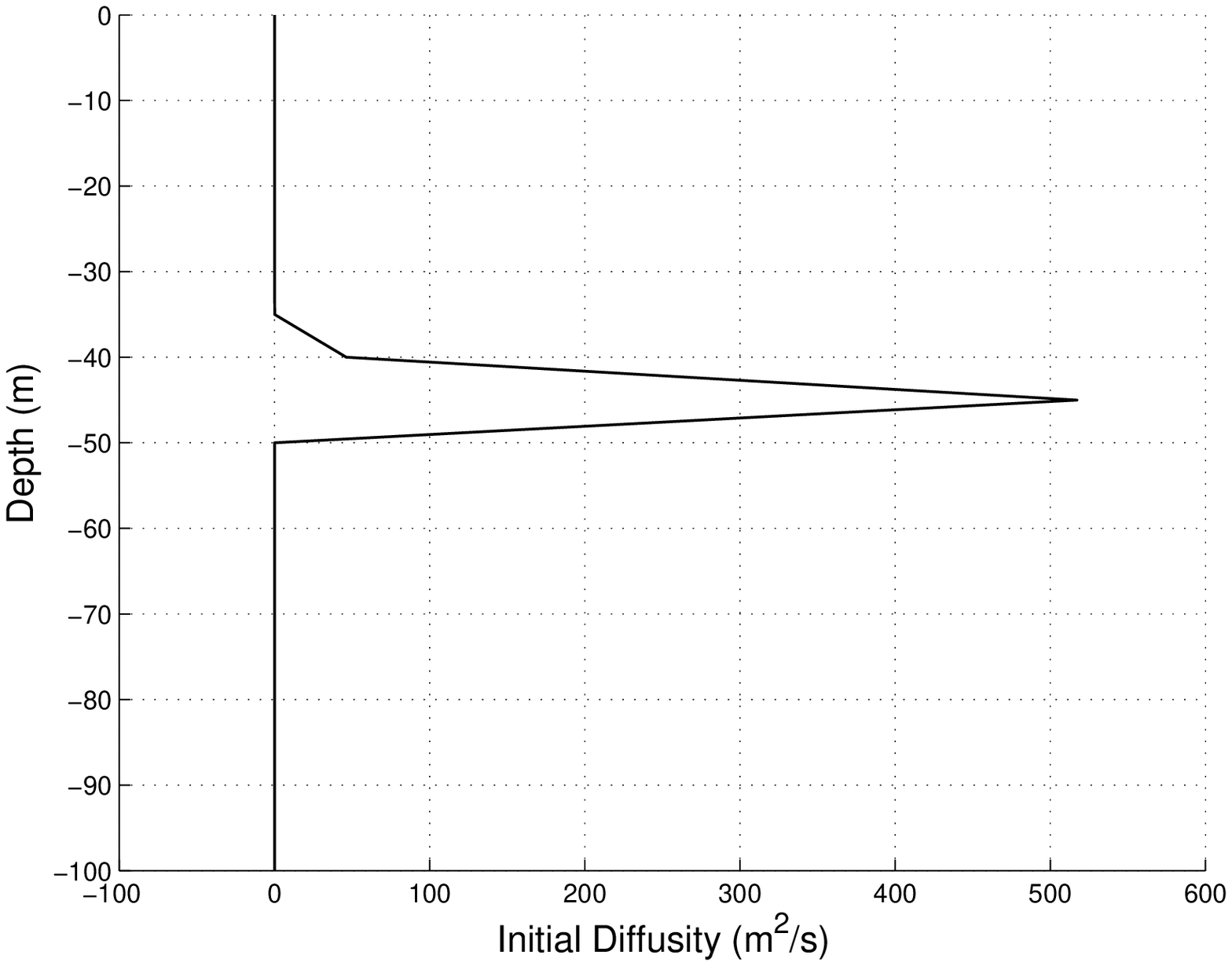}
\end{tabular}

\textbf{Figure 8}: Initial diffusivity for formulation R224.
\end{center}

\begin{center}
\begin{tabular}{ccc}
\hskip -1.5cm\includegraphics[scale=0.44]{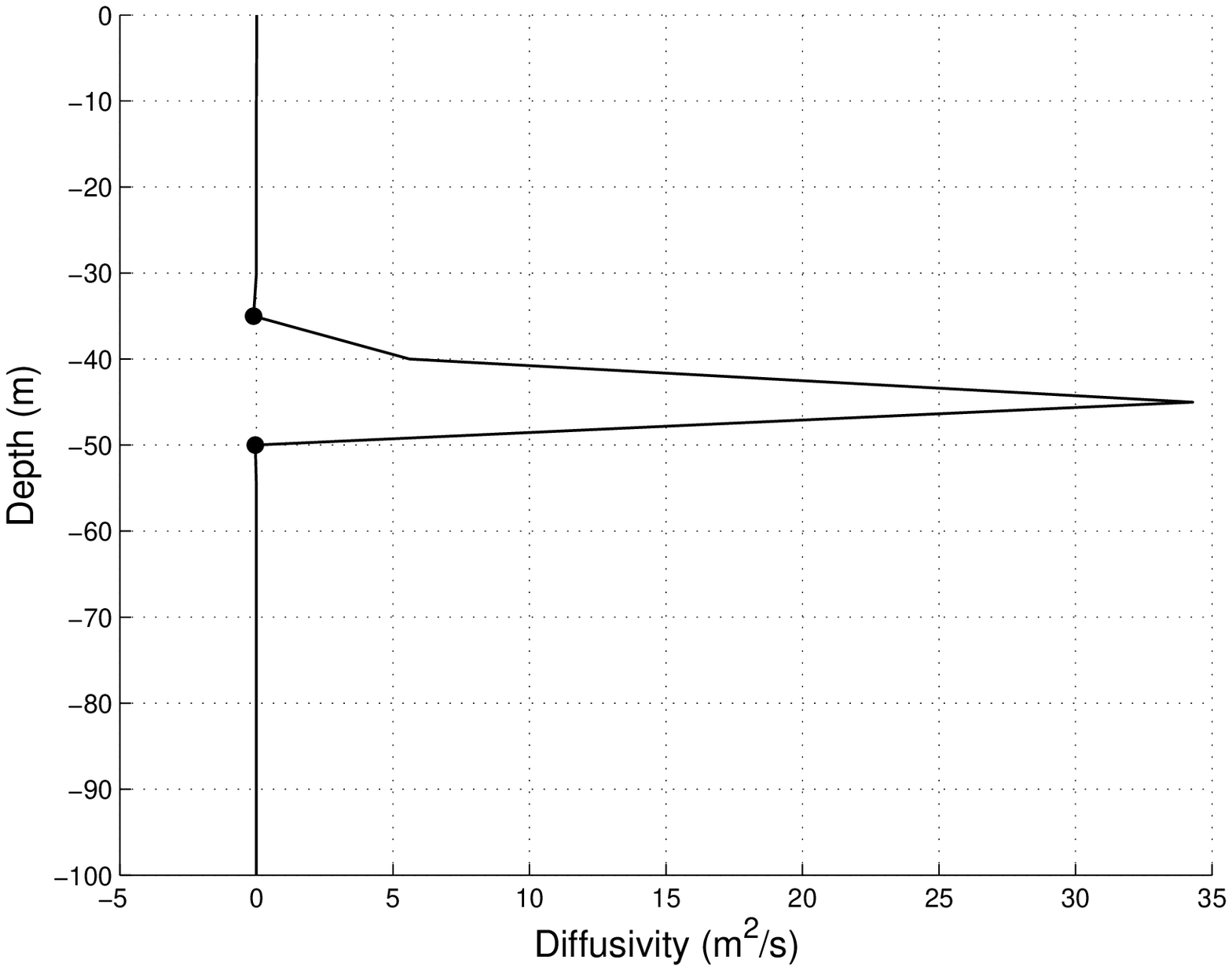} & &
\includegraphics[scale=0.44]{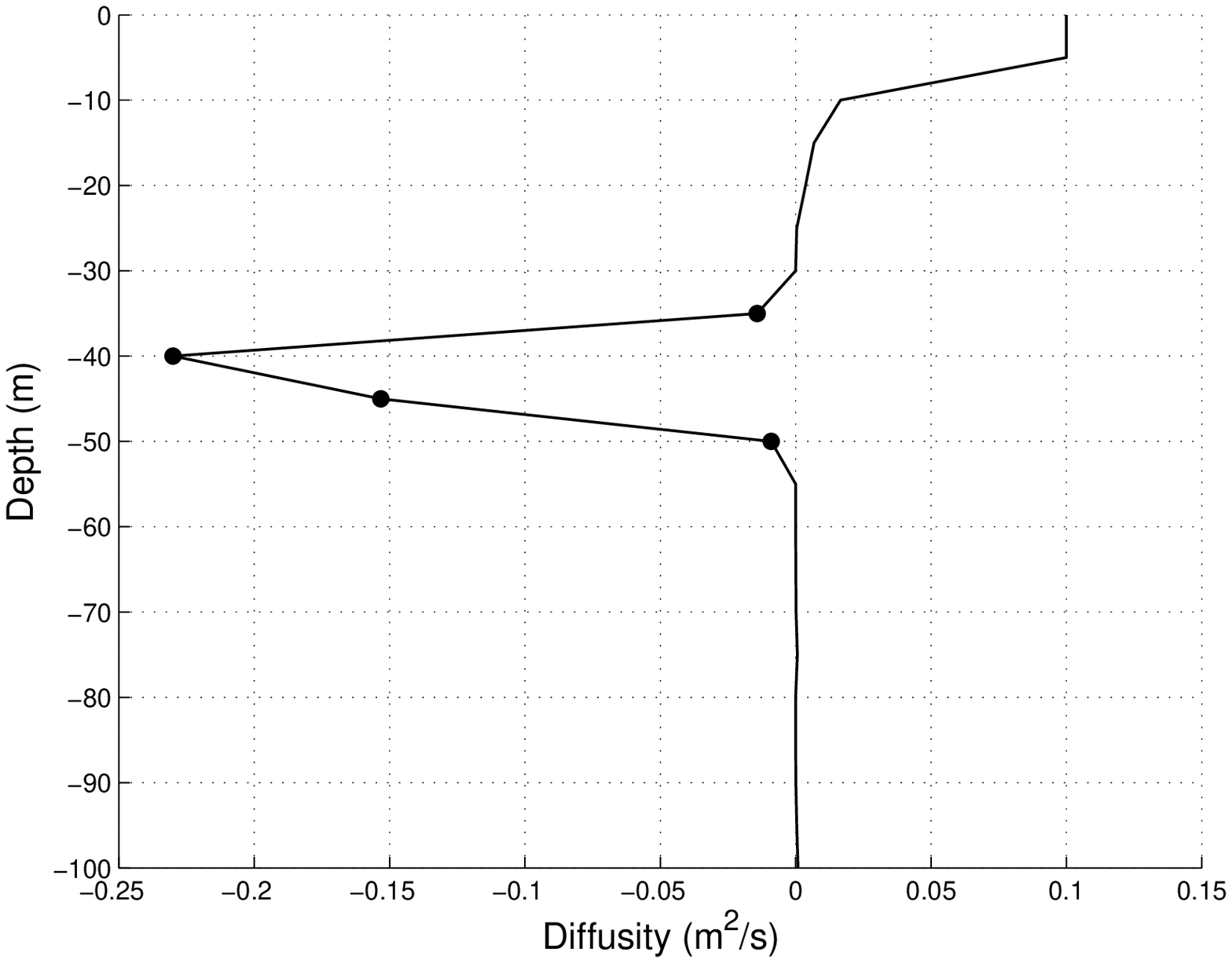} \\
\end{tabular}

\textbf{Figure 9}: Initial diffusivity for formulations R213 (left-hand side) and R23 (right-hand side).
The negative values are marked by a point.
\end{center}

\begin{center}

\includegraphics[scale=0.5]{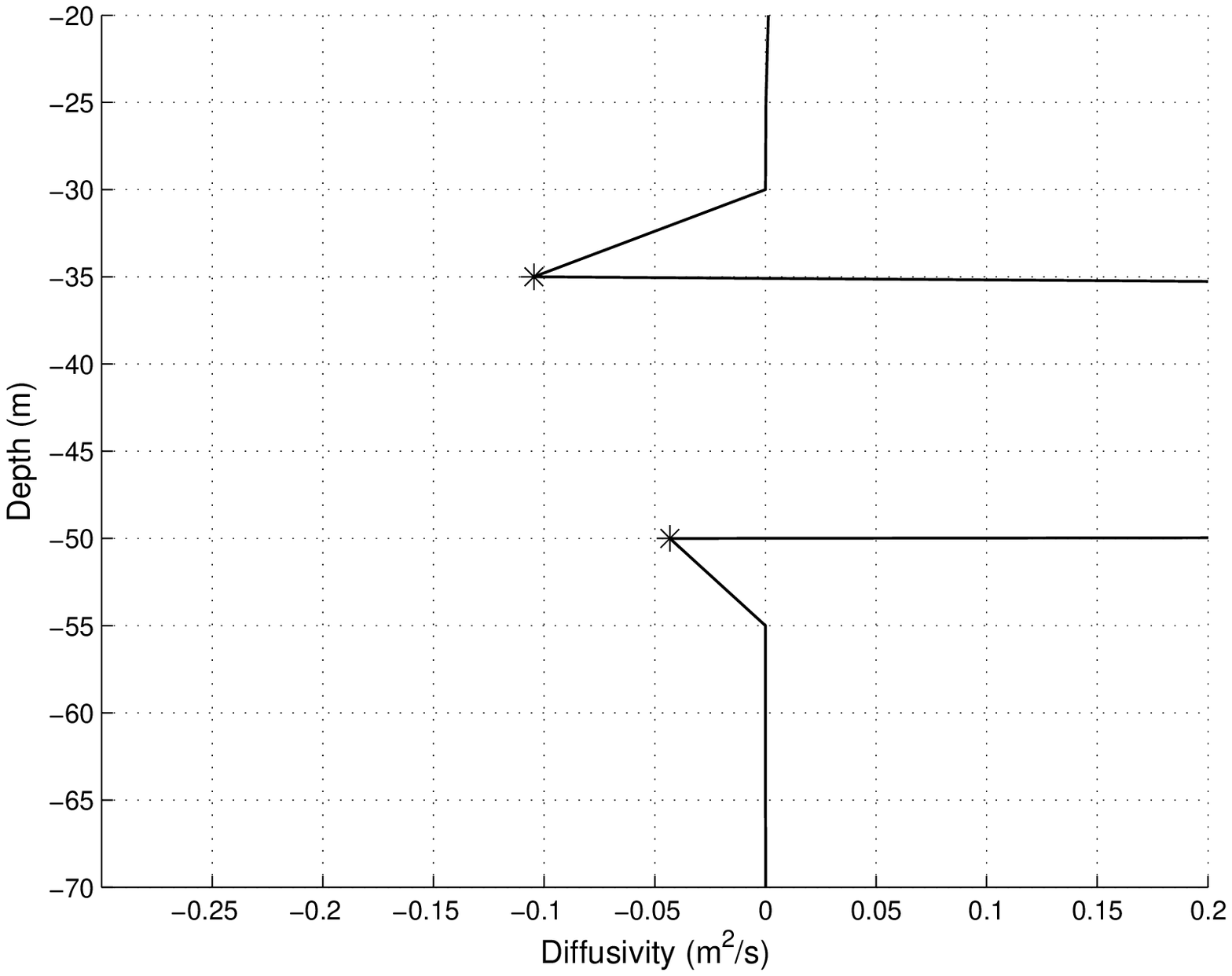}

\textbf{Figure 10}:
Zoom on the R213 negative values marked by an asterisk. \\

\end{center}

\medskip
The model is integrated for 48 hours. The grid spacing is equal to 5 meters and the
time step is equal to 60 seconds.
The zonal wind is equal to $11.7 \hskip 0.1 cm m.s^{-1}$ (eastward wind). 
The meridional wind
is equal to $0.4 \hskip 0.1 cm m.s^{-1}$ (northward wind). The
buoyancy flux is equal to $-1.10^{-6} \hskip 0.1 cm
kg.m^{-2}.s^{-1}$ ($\simeq -11W/m^2$). 

\medskip
The results are displayed on figure 11. The R224 model produces a
homogeneous seventy meters deep mixed layer. This layer is
homogeneous because we have applied a negative buoyancy flux at the surface which stabilizes the flow. 

\medskip
We notice an increase in the zonal surface current, in comparison with the initial surface current, which is in agreement with an eastward wind at the surface.
The northward wind at the surface is too weak to really modify the initial meridian surface current.

\hskip -2.3cm \includegraphics[scale=0.40]{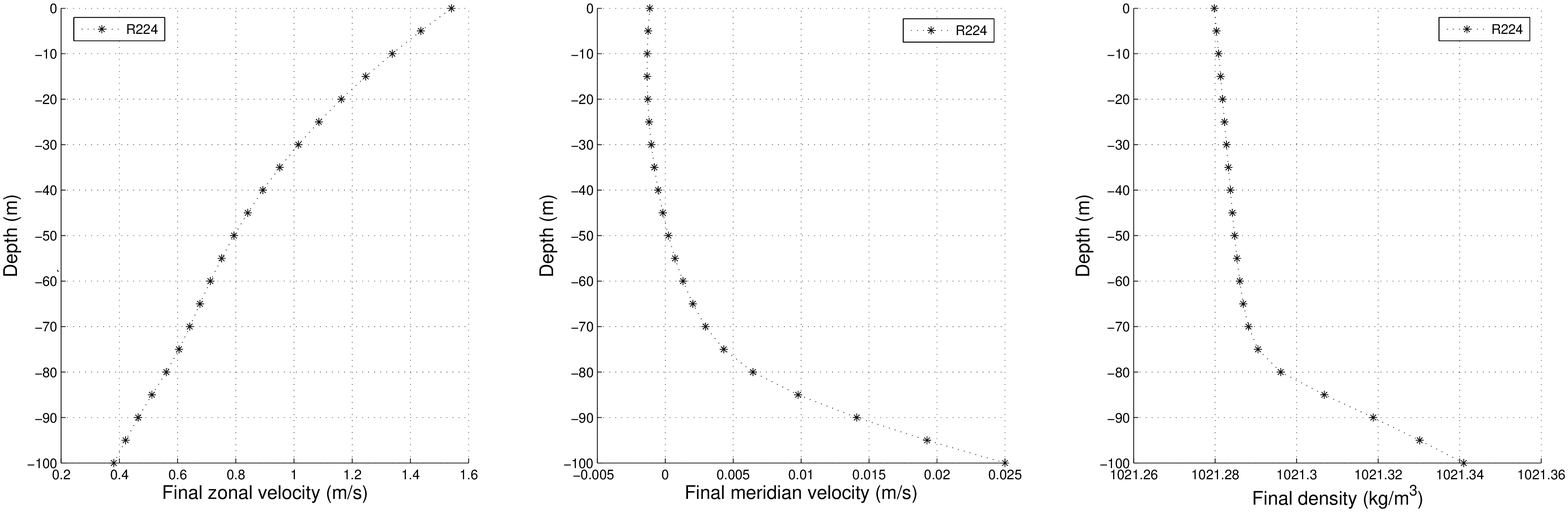}

\begin{center}
\textbf{Figure 11}: Zonal velocity profile (left
position), meridional velocity profile (medium position) and
density profile (right position) in case of the R224 formulation.
\end{center}

\hskip 3.2 cm\includegraphics[scale=0.44]{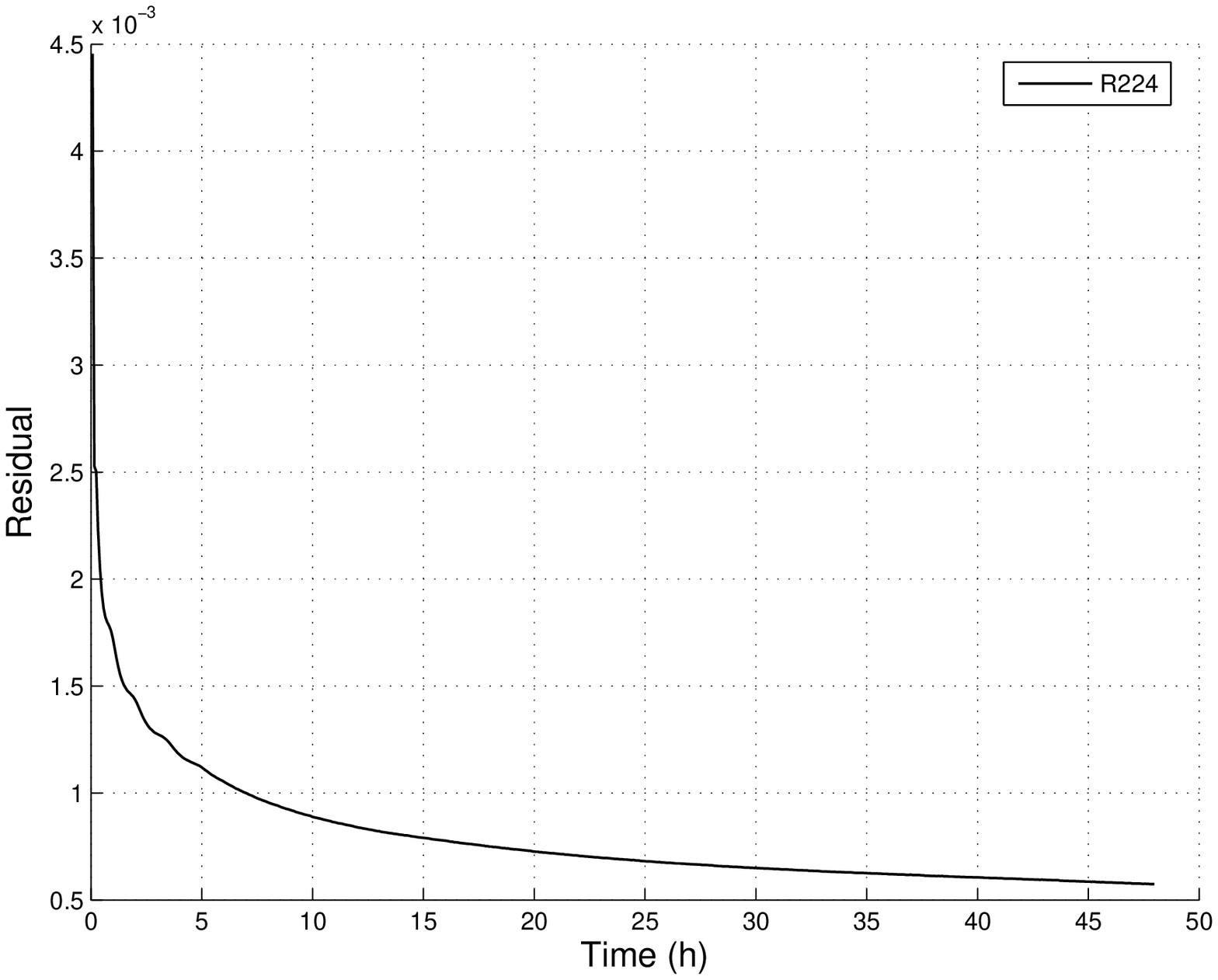}

\begin{center}
\textbf{Figure 12}: Evolution of residual values in case of the R224 formulation. 
\end{center}

On figure 12, the residual values display a good numerical convergence for R224
model (solid line).


\subsubsection{Case 3: A long time case}

We simulate a long time case. So, the
model is integrated for 10000 hours. The initial
profiles represent the ocean mean state on June 17, 1991. The
buoyancy flux is equal to $-1.10^{-6} \hskip 0.1 cm
kg.m^{-2}.s^{-1}$ ($\simeq -11 \hskip 0.1 cm W/m^{2}$).  The surface zonal wind
is equal to $5.4\hskip 0.1 cm m.s^{-1}$ (eastward wind) and the
surface meridional wind is equal to $0.9 \hskip 0.1 cm m.s^{-1}$
(northward wind).

\medskip
The initial zonal velocity profile (see Figure 13) displays two eastward currents located at the surface and around to $-70 \hskip 0.1 cm m$
as well as two westward currents located around to $-45 \hskip 0.1 cm m$ and $-90 \hskip 0.1 cm m$.
The initial meridian velocity profile (see Figure 13) displays several southward currents and northward currents.
The main southward current is located around to $-55\hskip 0.1 cm m$ 
while the main northward current are located around to $-90\hskip 0.1 cm m$ and at the surface.
Furthermore, on initial density profile (see Figure 13), we notice a thirty five meters deep mixed layer according to Peters and al's \cite{Pe89} criterion.

\hskip -2.3cm\includegraphics[scale=0.44]{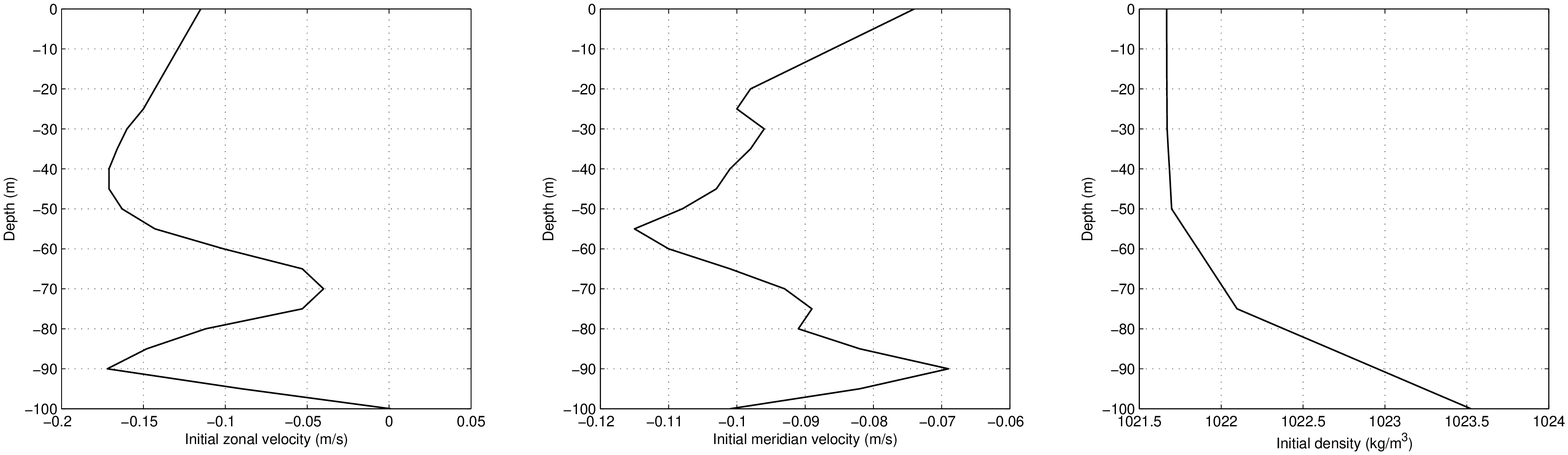}

\begin{center}
\textbf{Figure 13}: The initial profile of zonal velocity, meridional velocity, density (left to right).\\
\end{center}

The numerical results are displayed on figure 14. The three
turbulence models give a linear profile for the simulated time. This
fact corroborates the existence of a linear equilibrium solution
obtained by Bennis and al \cite{Be07}. 

\hskip -2.5cm \includegraphics[scale=0.43]{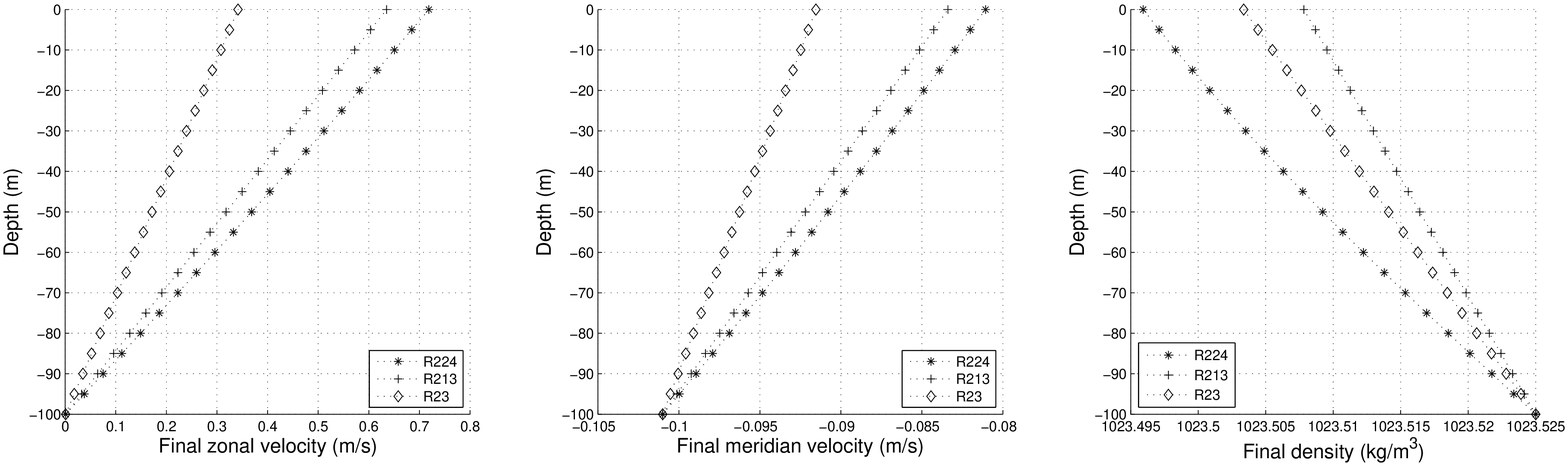}

\begin{center}
\textbf{Figure 14}: Comparison of zonal velocity profiles (left
position), meridional velocity profiles (medium position) and
density profiles (right position).
\end{center}

The residual values are displayed on figure 15. Notice that the numerical
convergence to steady states is good.

\begin{center}
\includegraphics[scale=0.45]{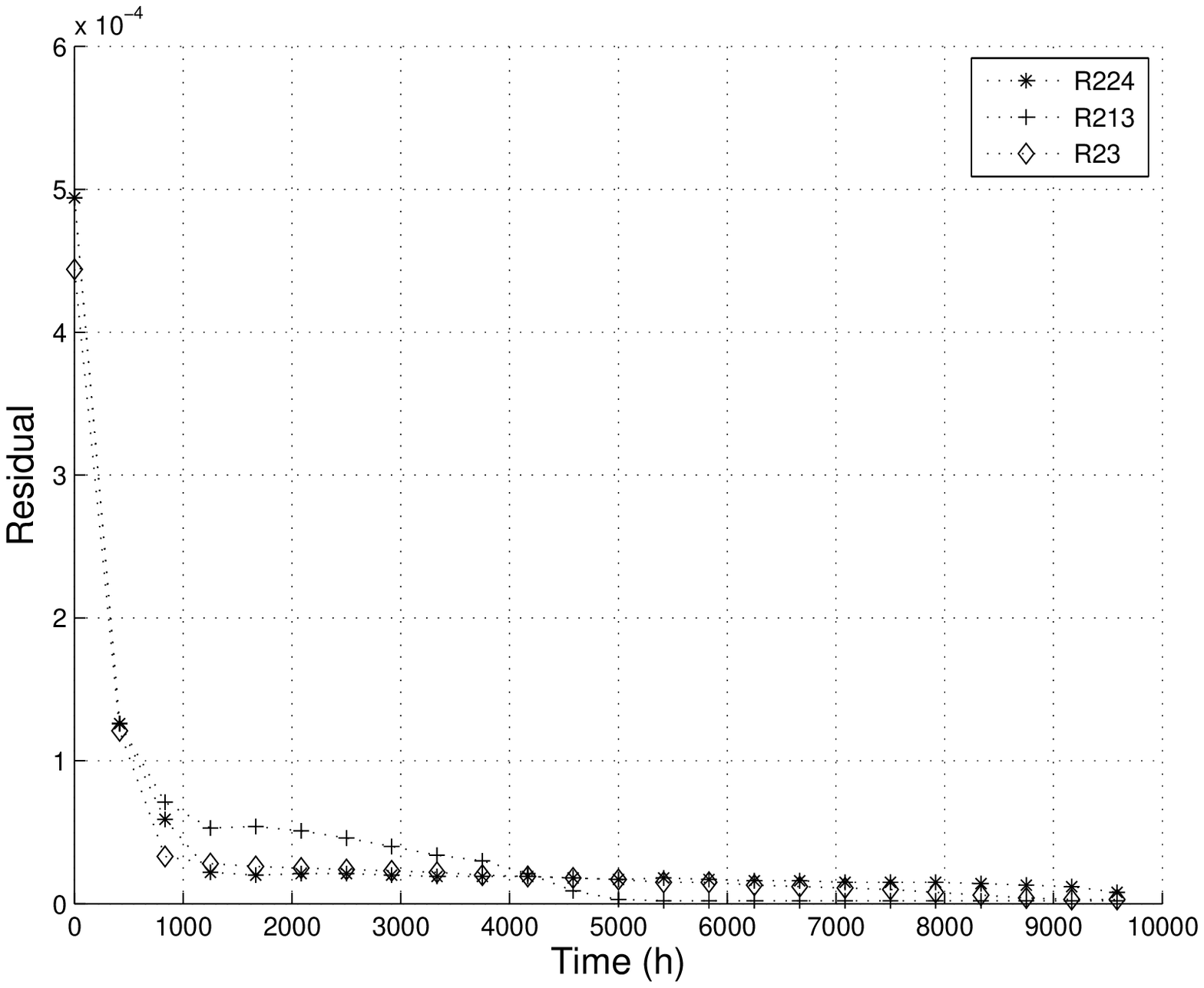}

\textbf{Figure 15}: Comparison of residual values. 
\end{center}

As the equilibrium solutions (zonal velocity $u^{e}$, meridian velocity $v^{e}$ and density $\rho^{e}$) are unique in case of the R224 model (see Bennis and al, 2007 \cite{Be07}), we can compare the theoretical solutions with the numerical solutions. We remember hereafter the theoretical equilibrium solutions which satisfy the following system: 
$$
\dfrac{\partial }{\partial z}\left( f_{1}\left( R^{e}\right) \dfrac{\partial u^{e}}{%
\partial z}\right) =0, \quad 
\dfrac{\partial }{\partial z}\left( f_{1}\left( R^{e}\right) \dfrac{\partial v^{e}}{%
\partial z}\right) =0,\quad 
\dfrac{\partial }{\partial z}\left( f_{2}\left( R^{e}\right) \dfrac{\partial
\rho^{e} }{\partial z}\right) =0.
$$ 

The  theoretical steady-state profiles for velocities and density are obtained by integrating the previous system with respect to z, taking account the boundary conditions at $z=-h$: 
$$
u^{e}\left( z\right) =u_{b}+\dfrac{V_{x}\rho_{a}}{\rho_{0}f_{1}\left( R^{e}\right) }\,\left(
z+h\right) ,\quad 
v^{e}\left( z\right) =v_{b}+\dfrac{V_{y}\rho_{a}}{\rho_{0}f_{1}\left( R^{e}\right) }\,\left(
z+h\right) ,\quad
\rho ^{e}(z)=\rho _{b}+\dfrac{Q}{f_{2}\left( R^{e}\right) }\,\left(
z+h\right) . 
$$

The equilibrium richardson number ($R^{e}$) can be interpreted as the intersection of the
curves $k\left( R\right) =\dfrac{\left( f_{1}\left( R\right) \right) ^{2}}{%
f_{2}\left( R\right) }$ and $h\left( R\right) =CR$ with $\displaystyle C=-\frac{\rho_{a}^{2}(V_{x}^{2}+V_{y}^{2})}{g Q \rho_{0}}%
$. In case of R224 model, the graph of function $k$ and $h$ for $Q<0$ is plotted on figure 16.

\begin{center}
\includegraphics[scale=0.42]{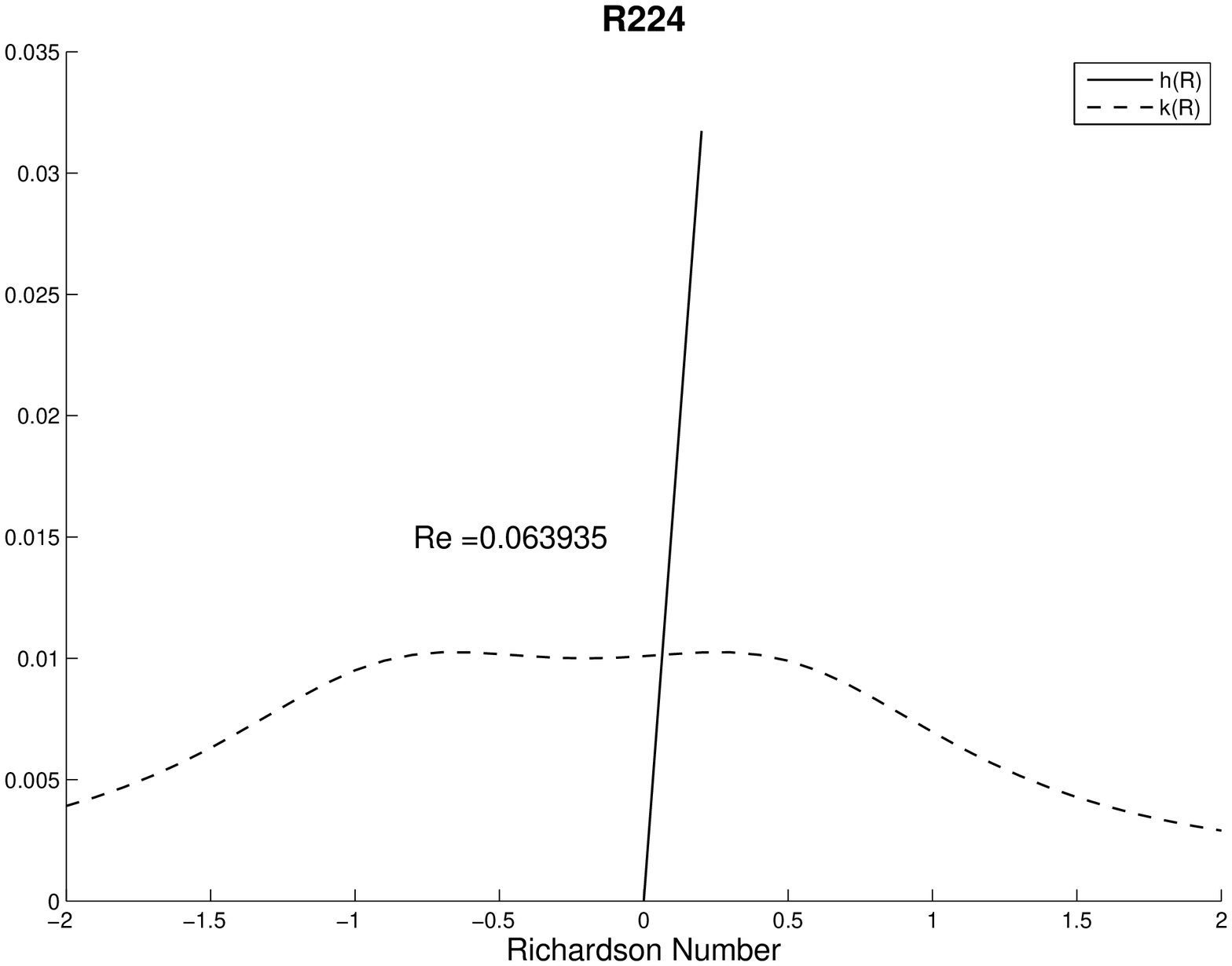}

\textbf{Figure 16}: Equilibrium richardson number - R224 model.
\end{center}

So, we compute the theoretical solutions with the equilibrium richardson number ($Re=0.063935$) given by the equation $h(R)=k(R)$. The theoretical and numerical equilibrium solutions are displayed on figure 17.\\

\hskip -2.5 cm\includegraphics[scale=0.39]{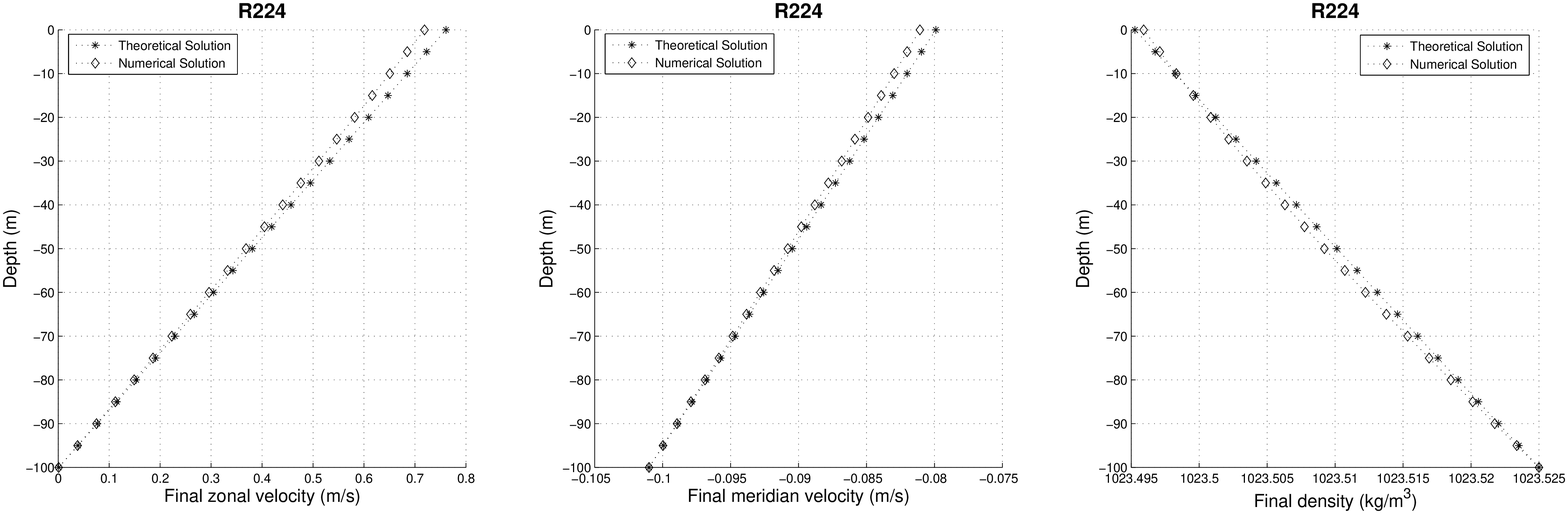}

\begin{center}
\textbf{Figure 17}: Theoretical and numerical equilibrium solutions - R224 model.
\end{center}

We obtain similar profiles and the small differences are due to numerical scheme which we use.

\section{Conclusion}

\subsection{Code validation}

We observe, in case 1, a mixed layer induced by the wind stress that is a physical phenomenon.
Moreover, we notice, in all cases, an increase in the zonal surface current when we apply an eastward wind ($u>0$) at the surface.
In the same order, an efficiency northward wind ($v>0$) at the surface causes an
increase in the meridional surface current. These results are in
agreement with the physical reality.
Furthermore, the residual values correctly converge to steady states, corresponding to rather high levels of turbulent diffusion. Theses previous
observations validate our code that gives a good representation of the
physical reality.

\subsection{Comparison of three turbulence models}

\medskip
Our comparison is based on three criteria: the mixed layer depth,
the surface current intensity and the pycnocline's form. The mixed
layer depth is obtained with a density difference criterion. This
difference is equal to $0.01 \hskip 0.1 cm kg.m^{-3}$. We use the
surface current values to determine the surface current intensity.
The density gradient is used to determine the pycnocline's form. We
summarize the results hereafter.

\begin{itemize}

\item[-] In the particular case 1, the R213, R23 and R224 models give a same mixed layer depth.
In terms of surface current, those simulated by R224 and R213 models are of the same kind.
However the R224 surface current is slightly stronger than the R213 surface current.
The R23 model underestimates this current. Furthermore, the pycnocline's form are similar
for R213, R23 and R224 models.

\item[-] In the particular case 2, the R213 and R23 models have negative diffusivities at the initial time.
Therefore, these models are not physically valid in this case. This
problem comes from static instability zone in the initial density profile. Hence, we can not use these models in this case.
Then, only R224 is valid. The R224 model produces a homogeneous mixed layer. So, 
we do not compare the pycnocline's form since we have only one mixed layer.

\item[-] Furthermore, the long time profiles are linear for all models that is in
agreement with Bennis and al \cite{Be07}. In case of R224 model, the theoretical and the numerical profiles 
are similar.

\end{itemize}

Globally, the R224 model has the same behavior as the Pacanowski
and Philander model (R213 model) and we can use it in more
situation. For example, R224 model can be used in almost all of convective cases.
Moreover, the R224 numerical equilibrium solution is in agreement with the R224 theoretical equilibrium solution.

\bigskip
\textbf{\large{Acknowlegments}}

The research of T. Chacon  Rebollo and Macarena Gomez Marmol has been partially funded by Spanish Government Grant MTM2006-01275. Furthermore, the authors express their gratitude to Pascale Delecluse, Assistant Director of Research at Meteo-France,
for her relevant advice.

%
%
\bibliographystyle{siam} \bibliography{Bib}

\end{document}